\documentclass[aps,pra,a4paper,superscriptaddress,longbibliography,reprint]{revtex4-1}

\usepackage[T1]{fontenc}
\usepackage{newtxmath,newtxtext}

\usepackage{amsmath}
\usepackage{amsfonts}
\usepackage{bm}

\usepackage{graphicx}

\usepackage{color}

\newcommand{\eq}[1]{Eq.~(\ref{#1})}
\newcommand{\fig}[1]{Fig.~\ref{#1}}

\newcommand{\etal}{\textit{et al.}}
\newcommand{\ofr}{(\textbf{r})}

\begin{document}

\title{Machine learning based modeling of disordered elemental semiconductors: understanding
the atomic structure of a-Si and a-C}

\author{Miguel A. Caro}
\email{mcaroba@gmail.com}
\homepage{http://miguelcaro.org}
\affiliation{Department of Chemistry and Materials Science,
Aalto University, 02150, Espoo, Finland}
\affiliation{Department of Electrical Engineering and Automation,
Aalto University, 02150, Espoo, Finland}

\date{\today}

\begin{abstract}
Disordered elemental semiconductors, most notably a-C and a-Si, are ubiquitous
in a myriad of different applications. These exploit their unique mechanical and
electronic properties. In the past couple of decades, density functional theory (DFT)
and other quantum mechanics-based computational simulation techniques have been successful
at delivering a detailed understanding of the atomic and electronic structure of
crystalline semiconductors. Unfortunately, the complex structure of disordered
semiconductors sets the time and length scales required for DFT simulation of
these materials out of reach. In recent years, machine learning (ML) approaches to atomistic
modeling have been developed that provide an accurate approximation of the DFT
potential energy surface for a small fraction of the computational time. These ML
approaches have now reached maturity and are starting to deliver the first conclusive
insights into some of the missing details surrounding the intricate atomic structure
of disordered semiconductors. In this Topical Review we give a brief introduction
to ML atomistic modeling and its application to amorphous semiconductors. We then
take a look at how ML simulations have been used to improve our current understanding
of the atomic structure of a-C and a-Si.
\end{abstract}

\maketitle

\section{Introduction}

Since the inception of the first experimental semiconductor diodes in the early
1900s the presence of semiconductors in daily appliances as well as high-tech
equipment has grown exponentially. Today, virtually all equipment incorporating electrical
circuits or electronic components, including computers and mobile phones, have parts
made of silicon. Commercially successful light-emitting diodes (LEDs) and laser
diodes (LDs) also use semiconductors, most often III-V compounds. Many of the familiar
applications of semiconductors use their crystalline forms, and the degree of
crystallinity often dictates the quality of the device. Indeed, in LEDs even
tiny amounts of crystallographic defects can severely deteriorate device
performance~\cite{krames_2007,humphreys_2008}.

On the other hand, amorphous semiconductors, notably a-C and a-Si, can have useful
properties of their own. Whether they offer actual performance improvements over
crystalline forms for specific applications, or a significantly cheaper and more
scalable fabrication process gives them a practical advantage, these materials
are widely used for applications where their electronic, chemical, mechanical and
optical properties are exploited. Hydrogenated a-Si (a-Si:H) is used to fabricate
low-cost solar cells~\cite{stuckelberger_2017}. More generally, a-Si and its
derivatives find uses in applications where a
cost-effective alternative to crystalline Si (c-Si) is desirable, or where less
stringent growth conditions (e.g., lower deposition temperature) are required~\cite{rech_1999}.
This includes such applications as thin-film transistors (TFTs)~\cite{schroder_1991},
liquid-crystal displays (LCDs)~\cite{chen_2005} and medical X-ray imaging~\cite{karim_2004}.
a-C is even more versatile than
a-Si since its properties can be more or less continuously tuned between those of
graphitic carbon (g-C) and diamondlike carbon (DLC)~\cite{robertson_2002}. Current uses of a-C
and a-C thin films include biocompatible and bioimplantable devices (such as hip
replacement implants)~\cite{tiainen_2001}, electrochemical sensors for in-vivo
analysis~\cite{laurila_2017}
and hard coatings for tribological applications~\cite{donnet_2007}. Furthermore, modified a-C
such as oxygen-rich a-C (a-COx)~\cite{santini_2015}, nitrogen-doped a-C (a-C:N)~\cite{etula_2021},
different carbon hybrid materials~\cite{laurila_2017}, nanocarbons modified under extreme
conditions~\cite{shiell_2018,shang_2021,sundqvist_2021} and the wider family of disordered carbons
are starting or expected to make their way to emerging
applications in energy storage~\cite{wang_2021}.
More generally speaking, carbon-based materials are envisioned to be key in the
transition to renewable raw materials utilization and the bioeconomy~\cite{arasto_2021}.

Unsurprisingly, the diversity and complexity of the atomic structure of a-C and a-Si
pose serious challenges for experimental characterization. For crystalline materials,
common structural characterization methods, like X-ray diffraction (XRD), rely on
the \textit{periodic} structure of crystals, and are thus less useful to characterize amorphous
materials. Instead, the structure of a-C and a-Si (and other disordered materials)
can be characterized using experimental techniques such as X-ray photoelectron or
absorption spectroscopy (XPS and XAS, respectively), Raman spectroscopy and neutron
scattering. A very complete summary of experimental structure characterization
techniques for a-C has been given by Robertson~\cite{robertson_2002} (these techniques
are also relevant for the characterization of a-Si).
In our strive to understand the atomic structure of disordered materials, computational
atomistic modeling techniques arise as an obvious choice: by being able to model the
interatomic energies and forces between atoms, and update or optimize their positions
accordingly, we can effectively ``look'' at the atomic structure. To access the length
and time scales involved in modeling amorphous materials accurately, machine-learning
interatomic potentials (MLPs) have emerged in recent years as game changers in the
field~\cite{deringer_2019}.

In this Topical Review we will first discuss general considerations pertaining to atomistic
modeling of amorphous semiconductors. We will then give a brief introduction
to MLPs that should be
accessible to those with basic understanding of atomistic simulations, either coming
from a (modest) density-functional theory (DFT) or classical molecular dynamics (MD)
background. We will then show how MLPs have enabled
a new degree of realism in modeling a-Si and a-C, arguably the two most important
elemental amorphous semiconductors. We will end with a brief discussion of the state of
the field and an outlook for the future.

\section{$\text{a-C}$ and $\text{a-Si}$ atomistic simulation}

The main fundamental difference between a crystalline and an amorphous semiconductor is
the lack of long-range atomic order in the latter. The other differences
(electronic and thermal conductivities, electronic and optical band gap, mechanical
properties, etc.) ultimately stem from the differences in the atomic structure. In a-Si,
local atomic structures are usually 4-fold tetrahedral motifs due to $sp^3$ chemical
bond hybridization. Lower (3-fold) and higher (5-fold) coordinations in a-Si are typically
considered coordination defects~\cite{pantelides_1986}. Thus, the structural complexity
in a-Si is compounded by the
interplay between the local arrangement of nearby stable 4-fold motifs and the existence
of coordination defects in the amorphous network. In the case of a-C the situation is
significantly more complex since stable chemical motifs in elemental carbon can be due
to $sp$ (2-fold), $sp^2$ (3-fold, ``graphite-like'') and $sp^3$ (4-fold, ``diamond-like'')
hybridizations. The atomic structure of a-C is consequently diverse, making a-C effectively
a \textit{range} of materials, rather than just a material, typically characterized to a first
approximation by the relative amount of $sp^2$ and $sp^3$ carbon. The $sp^2$-rich forms
of a-C are low in mass density (down to 2~g/cm$^3$ and less~\cite{robertson_2002}),
whereas the $sp^3$-rich
forms have high mass density and are often referred to as ``diamond-like'' or ``tetrahedral''
a-C (DLC and ta-C, respectively)~\cite{robertson_2002}.
To complicate things, a-C and a-Si can exist with different
degrees of hydrogenation, where some of the C-C and Si-Si bonds are replaced by C-H and
Si-H bonds. These materials are usually referred to as a-C:H and a-Si:H, respectively, and their
properties, especially mass density, may differ from those of the hydrogen-free forms. We
note here in passing that pure a-C and a-Si do not exist in practice, and some level
of impurities, mostly H and O, are always present in experimental
samples~\cite{carlson_1977,robertson_2002}.

The standard
for predicting the structure of materials at the atomic scale is density functional
theory (DFT). DFT is a quantum mechanical method, providing an approximation to the
Schr\"odinger equation. Its popularity stems from the computational efficiency of the
Kohn-Sham formulation of DFT~\cite{hohenberg_1964,kohn_1965,martin_2004}, which resides
at a ``sweet spot'' of accuracy vs CPU
cost. DFT is routinely used to study crystals and to carry out crystal structure
prediction~\cite{oganov_2006,pickard_2011}, benefiting from the fact that crystals can be
represented with small primitive unit cells, often comprising just a handful of atoms.
Unfortunately, even DFT can become prohibitively expensive to model amorphous materials,
which lack short-range order. In practice, DFT has been used to study amorphous materials
in a limited way by employing the ``supercell'' approach. A supercell is made of tens
or, at most, a few hundreds of atoms in periodic boundary
conditions~\cite{marks_1996,marks_2002,caro_2014}.
Thus, effectively, amorphous compounds are modeled as crystals with very large unit
cells.

The accuracy of the supercell approach to model real amorphous materials improves
with system size, but not only. To provide a realistic view of an amorphous structure
it is necessary to collect statistics via configurational sampling, since each individual
supercell will, in general, look different from another. More critically, while a
``single-point'' DFT calculation (i.e., a calculation where the atomic positions are
not updated) for a given structure may be affordable even for
relatively large system
sizes of a couple or few thousands of atoms, placing the atoms in configurations
that resemble real amorphous structures is far from trivial~\cite{marks_2002}.

Computational structure-generation protocols for amorphous materials come in two
flavors. On the one hand, there are direct protocols trying to mimic the experimental
growth process as closely as possible. This is for instance the case for simulated
deposition, where the attachment of atoms onto a growing film is simulated one atom
at a time~\cite{kaukonen_1992,marks_2005,caro_2018,caro_2020c}.

On the other hand, there exist indirect protocols that rely on initially randomizing
the atomic positions and subsequently updating these positions. The positions can be
updated by either ``relaxing'' the structure (e.g., using gradient-descent optimization
along the direction of the forces)~\cite{galli_1989,caro_2014} or by carrying out
molecular dynamics (MD) with a
rapidly decreasing temperature profile (``quench''
simulations)~\cite{marks_2002,detomas_2016,deringer_2017,wang_2022}.
Sampling protocols
designed for free-energy sampling at given thermodynamic conditions~\cite{christ_2010}
are often not
a good choice to generate amorphous structures, since 1) amorphous materials are
usually \textit{metastable} and 2) these free-energy sampling methods can be prohibitively
expensive because they rely on many individual evaluations of the potential energy surface
(PES). A relatively
recent comparison between generation methods for a-C modeling, but lacking some of
the latest developments with machine-learning (ML) interatomic
potentials~\cite{caro_2018,caro_2020c}, has been given in Ref.~\cite{laurila_2017}.

Arguably, the most popular protocol to generate atomistic amorphous structures is
the MD-based ``melt-quench'' protocol~\cite{marks_2002}, which resembles how glass is made
in reality~\cite{shelby_2020}. In a melt-quench simulation a material is first heated to high temperature
$T$ until it melts. This liquid sample is then kept at high temperature to ensure a
disordered but relatively low-energy distribution of atoms (e.g., much lower in energy
than random) to provide a reasonable starting point. Then, the liquid is quickly
quenched down to a temperature well below the solidification temperature. Since the
process is so fast, the different atomic motifs are trapped into local minima, giving
an amorphous structure as a result. How fast the system is cooled down (the quench rate)
will determine the quality of the structure. A very fast quench will lead to numerous
defects, for example under (3-fold) and over (5-fold) coordinated atoms in a-Si~\cite{deringer_2018b}.
A very slow quench will (theoretically) lead to formation of the thermodynamically stable
allotrope of the material, for example diamond-structure silicon. Besides a temperature
profile, imposed in MD through the use of a thermostat~\cite{berendsen_1984}, one may also couple the
simulation to a barostat~\cite{martyna_1994}, to control the pressure $P$. This enables exploration of
phases and phase transformations within widely varying thermodynamic conditions,
including some extreme conditions not accessible experimentally~\cite{muhli_2021b}.

\begin{figure}
    \centering
    \includegraphics{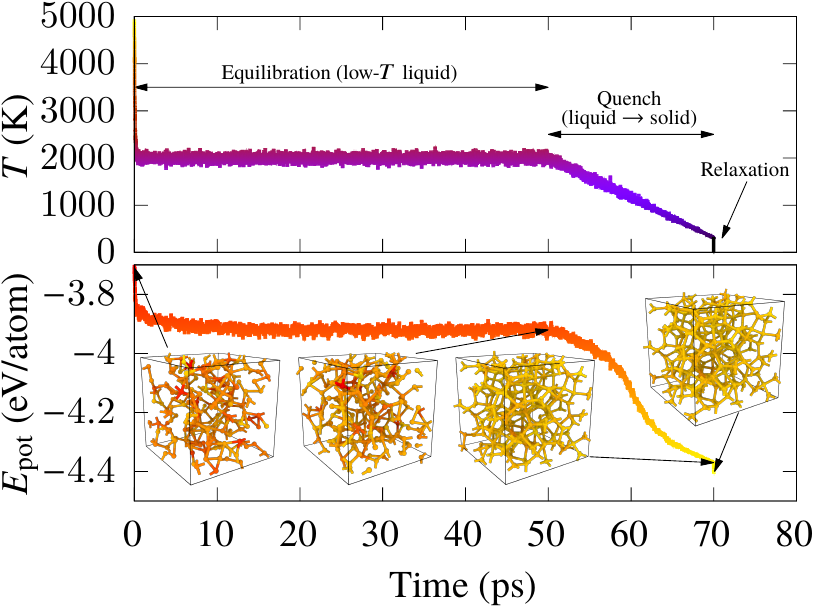}
    \caption{Melt-quench simulations of a-Si formation carried out with a GAP~\cite{bartok_2010}
    potential refitted~\cite{caro_2021b} from the database developed by Bart\'ok
    \etal~\cite{bartok_2018}. The simulations were carried out with the TurboGAP
    code~\cite{ref_turbogap,caro_2019}.}
    \label{01}
\end{figure}

Additional steps can be added before or after the quench, typically some sort of
annealing step. For instance, a carbon sample can be held at around 3500~K for a while
before quenching to favor graphitization~\cite{detomas_2016,detomas_2019,wang_2022}.
Or an a-Si sample can be annealed at a temperature below solidification (but still
significantly higher than room temperature) to heal defects~\cite{deringer_2018b}.

The melt-quench process leading to generation of a computational atomic structure is
exemplified for a-Si in \fig{01}. Initially, the sample, containing 216 atoms, has been heated
to a very high temperature of 5000~K to properly randomize the atomic positions. The
temperature is rapidly brought down to 2000~K, slightly above the melting temperature of
silicon, and kept there for some time (50~ps in our example). This is the equilibration
stage, where we aim to homogeneously distribute the available kinetic energy among all the
degrees of freedom and find local structures which are low in energy (for the given values
of $T$ and $P$). After equilibration, we quench the system down to 300~K using a linear
temperature profile. The evolution of the potential energy does not follow this linear
trend. Instead, there is a slow initial decrease in potential energy because the temperature
is too high to create stable motifs. This is followed by an accelerated decrease in energy
where these stable atomic motifs, tetrahedra in a-Si, are created at a fast pace. The final
stage of the quench corresponds to slow further decrease in potential energy, because of either
of two reasons: 1) all the Si atoms are already part of local tetrahedra or 2) there is not
enough kinetic energy to overcome local potential energy barriers and the atoms are trapped
into their local metastable structures. The actual situation is a combination of both factors.
Recall that, according to the virial theorem, as we linearly decrease the kinetic energy
there should be a corresponding linear decrease in potential energy, \textit{assuming that the
details of the potential energy surface do not change}. Therefore, the non-linear profile
observed in our example is indeed associated to the phase transition from the disordered
liquid to the amorphous solid.

After the MD quench, we further relax the structure using a static relaxation of the atomic
positions, following a gradient-descent minimization of the potential energy.
In \fig{01} we have additionally color-coded the Si atoms according to their local energy,
which can be extracted from a simulation with MLPs, as we detail in Sec.~\ref{02}. The
curious reader is encouraged to visit the \texttt{turbogap.fi} website for a series of tutorials
on how to run this type of simulation for a-Si and a-C.

Melt-quench simulations are popular because they provide a good compromise between
CPU cost and the quality of the generated structures. However, they do not (typically)
reproduce the experimental growth/formation protocol of the real material, and that
can have a non-negligible effect on the resulting atomic structure, as is for instance
the case for a-C~\cite{laurila_2017}. Unfortunately,
direct simulation protocols, such as deposition in a-C~\cite{marks_2005,caro_2018,caro_2020c},
are orders of magnitude more expensive than indirect methods.

Traditionally, direct simulation has been limited to empirical
interatomic potentials~\cite{biswas_1988,luedtke_1989,ramalingam_2001,marks_2005}.
These are very efficient computationally, relying on simple
mathematical functions that depend on the interatomic distances and angles and are
parametrized by fitting to experimental or first-principles data.
Popular examples of these potentials, which can often be used for both
Si and C (and even SiC) by adjusting the model's parametrization,
are Tersoff~\cite{tersoff_1988}, Stillinger-Weber~\cite{stillinger_1985,vink_2001},
EDIP~\cite{bazant_1997} (and its carbon version C-EDIP~\cite{marks_2000}),
REBO~\cite{brenner_1990,brenner_2002} and
ReaxFF~\cite{senftle_2016,vanduin_2003,srinivasan_2015,smith_2017}. However, these empirical
potentials lack the accuracy of DFT, and thus provide a representation of the potential
energy surface (PES)
of very inconsistent quality~\cite{behler_2007,detomas_2016,bartok_2018,detomas_2019,caro_2020c}.
While low-lying harmonic regions of the PES, i.e.,
the atomic configurations about equilibrium, can be reproduced with reasonable
accuracy, chemical reactions are described very poorly. Therefore, we find ourselves at an
impasse: on the one hand, the breaking and
formation of chemical bonds, critical to understand the growth of amorphous
materials, are not correctly described with affordable empirical potentials. On the other
hand, DFT can describe chemical reactions accurately but is computationally
unaffordable.

Fortunately, new atomistic simulation techniques based on machine learning (ML)
have emerged in recent years~\cite{behler_2007,bartok_2010,schutt_2020} that
bridge this huge gap in atomistic modeling of amorphous
materials~\cite{deringer_2017,bartok_2018}. These MLPs
rely on non-parametric fits to a reference PES, typically computed at the DFT
level of theory~\cite{behler_2017,deringer_2021}. While still significantly more
expensive than empirical force fields,
MLPs offer accuracy close to that of DFT for a tiny fraction of the CPU cost. MLPs
have had, in just the last few years, a huge impact on atomistic modeling of amorphous
and disordered materials, granting us atomistic insight into problems that were
completely out of reach less than a decade ago.

\section{ML interatomic potentials}\label{02}

\begin{figure*}
    \centering
    \includegraphics[width=\textwidth]{{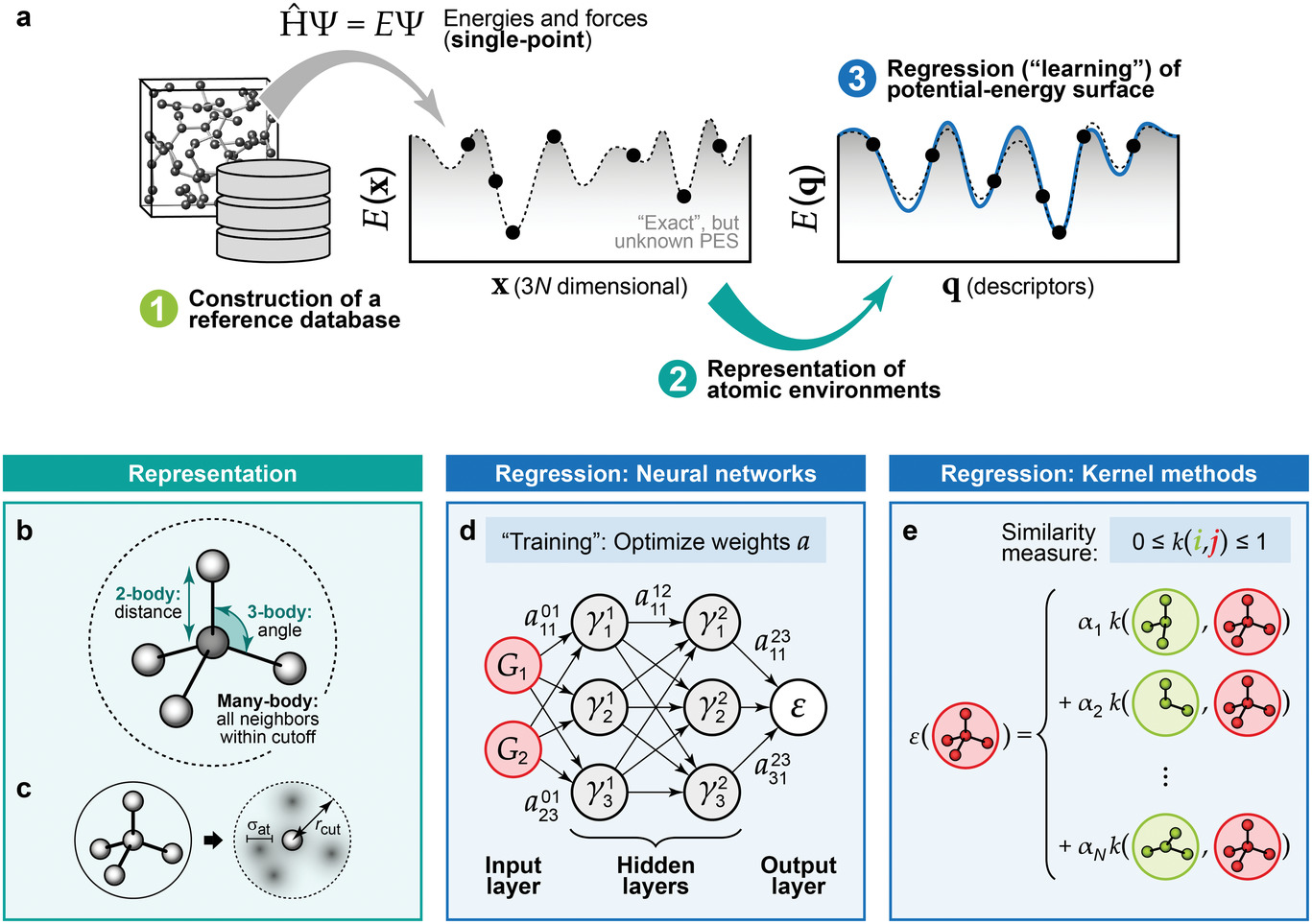}}
    \caption{a) Workflow of MLP training: a database of atomic structures and observables (energies,
    forces) is constructed, from which an ML algorithm is used to learn the PES as a function
    of atomic descriptors. b,c) Different kinds of descriptors commonly used to represent the
    atomic environments. d) Schematics of neural network prediction. e) Schematics of kernel-based
    prediction. Reprinted from Ref.~\cite{deringer_2019} with permission. Copyright
    (c) 2019 Wiley.}
    \label{04}
\end{figure*}

In this section we explain the whole MLP workflow, graphically summarized in \fig{04}. We
start with a brief general introduction to different popular ways
to represent the PES, with an emphasis on DFT. We will then give an overview of the
two main methodologies for learning and interpolating the DFT-PES based on a)
artificial neural networks (ANNs) and b) the related kernel-ridge regression (KRR)
and Gaussian process regression (GPR) methods. To take a
pedagogical approach, the introduction of these methodologies
will be preceded by a general introduction to relevant ML concepts (databases and
descriptors/features). We will compare ML potentials to DFT, on the one
hand, and classical force fields, on the other, to get an idea of what is possible now in
materials modeling, thanks to the introduction of MLPs, that was not possible just a few
years ago. For more comprehensive information, the
reader is referred to a recent book which nicely summarizes the current state of the
field~\cite{schutt_2020} including a chapter on GPR~\cite{csanyi_2020} and another
on ANNs~\cite{hellstrom_2020}, and to several overview
papers~\cite{behler_2017,bartok_2015,lilienfeld_2018,shapeev_2016,deringer_2019,deringer_2021}.

\subsection{Database construction (structure selection)}

\begin{figure}
    \centering
    \includegraphics[width=\columnwidth]{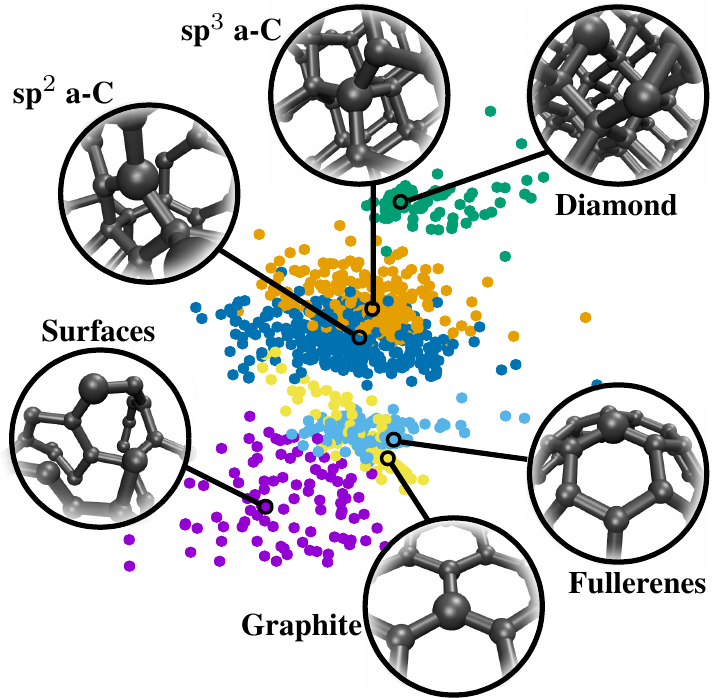}
    \caption{Low-dimensional embedding of high-dimensional data, used in this case to
    visualize the atomic structural diversity in a database constructed to fit an MLP for
    carbon. The closer two points are on the graph, the more closely the corresponding
    atomic environments resemble each other.
    Reprinted from Ref.~\cite{muhli_2021b} with permission. Copyright
    (c) 2021 American Physical Society.}
    \label{06}
\end{figure}

The creation of a new MLP starts with the generation of training data (\fig{04}a, step 1).
Many considerations
need to go into carefully crafting a suitable database for the problem at hand. There are
two main classes of MLPs depending on their scope: general- and single-purpose MLPs.
A \textbf{single-purpose} MLP is created with a very specific application in mind. In this
case, the MLP will be expected to perform with excellent (or even exquisite) accuracy for
the problem of interest, but there is no guarantee that it will perform even reasonably for any
other application. Recall that the MLP does not ``know'' about physics, chemistry, or the
Schr\"odinger equation; it only knows about data. Therefore, an MLP will only be able to
chart the portion of configuration space corresponding to the data that it was fed. Indeed,
single-purpose MLPs (and poorly designed general-purpose MLPs) will tend to ``blow up''
(MD jargon for when a force field becomes catastrophically unstable) when
tested on a problem for which they were not trained. A good example of a single-purpose MLP
would be one trained to reproduce the phonon dispersion curves of a crystalline material, for instance to
be used in thermal transport or thermal expansion calculations of c-Si and diamond/graphite.
A database suitable to fit
a good phonon MLP would typically incorporate many DFT calculations of structures distorted
from the equilibrium ones, by adding either homogeneously spaced or random strain transformations
to the unit cell in addition to rattling the atoms about their equilibrium positions. Furthermore,
the unit cells should span from the primitive unit cell up to larger unit cells which would
allow to capture interactions between distant atoms. An example of a single-purpose
MLP is the graphene GAP of Rowe \etal~\cite{rowe_2018}.

A \textbf{general purpose} MLP, on the other hand, is expected to perform reasonably accurately
in as many regions of configuration space as possible, and be resistant to blowing up. A good
general-purpose MLP is often very difficult to achieve because it may require prior
knowledge about which these regions are, and its training is consequently difficult to automate.
For instance, if one wants to fit an MLP to study the atomic structure of a-C surfaces, which
can be prohibitively expensive to generate with DFT using a melt-quench simulation (cf. \fig{01}),
how are sample surfaces sourced for the (single-point) DFT calculations that will serve as
reference for the MLP? In these cases,
\textit{iterative training}~\cite{deringer_2017,bernstein_2019} can help in improving
the accuracy of the MLP in regions of interest in configuration space and also to get rid
of pathological behavior. In our a-C surface example, iterative training would consist of
generating surface structures with an interim (low-quality) version of the MLP via melt-quench
simulations. Single-point DFT calculations are then performed on the final structures and this
data added to the training set. A new interim version of the MLP is trained and the whole
procedure is repeated until the MLP errors (compared to the DFT calculations) are below an
acceptable threshold. Besides regular sampling of known crystal phases, iterative training
can also be combined with less directed exploration of configuration space, such as
random structure search (RSS)~\cite{pickard_2011,pickard_2022}.

Finally, we can combine the features of a general-purpose MLP with those of a single-purpose
MLP, to improve the accuracy of the general-purpose MLP for specific applications, as has been
done for phonons in Si~\cite{george_2020} or fullerenes in C~\cite{muhli_2021b}.

A way to visualize these structural databases is via so-called
structure maps~\cite{de_2016,cheng_2020}. In these,
the similarities between different entries in
a database, i.e., between different atomic structures, can be plotted on a two-dimensional
map using low-dimensional embedding techniques~\cite{de_2016,cheng_2020}.
An example for the fullerene-augmented C MLP
mentioned earlier~\cite{muhli_2021b} is shown in \fig{06}. In this graph, each data point
represents an atom-centered environment and similar structures are \textit{clustered}
together. There is a transition from diamond structures to amorphous $sp^3$, then to
amorphous $sp^2$ and finally to different graphitic structures, including fullerenes.
These sketchmaps are a useful tool to glimpse at the composition of an entire database
and understand the relationships between the different structures.

\subsection{Reference representations of the potential energy surface}

A central objective of computational atomistic modeling is to gain access to an accurate
representation of the Born-Oppenheimer (BO) PES of a given
ensemble of interacting atoms. The BO approximation relies on decoupling
the electronic and nuclear degrees of freedom. That is, the BO-PES gives the total cohesive
energy of a set of interacting atoms as the \textit{electronic} ground state (GS) for
fixed nuclear positions~\cite{martin_2004}.
This approximation is valid in many situations, in particular
in condensed-matter physics, because of the mass difference between electrons and nuclei.
The electronic degrees of freedom evolve within much shorter time scales than the nuclear
degrees of freedom, and the atomic trajectories can be propagated in time treating the
nuclei as classical particles, following Newton's second law. The most popular approximation
used today to calculate the BO-PES is DFT~\cite{hohenberg_1964,kohn_1965,martin_2004}.
The fundamental tenet of DFT is that the total (cohesive) energy of a system of interacting
electrons in a external potential (given by the positively charged nuclei) is given by
a universal functional $E[n]$ of the electron density $n \ofr$, where the density that
minimizes the functional is the GS density and the energy of the GS is given by the
energy functional evaluated at the GS density~\cite{hohenberg_1964}:
\begin{align}
n_\text{GS} \ofr = \underset{n \ofr}{\text{argmin}} \, E [n \ofr], \qquad
E_\text{GS} = E [n_\text{GS} \ofr].
\label{03}
\end{align}
The practical means for solving \eq{03} are provided by the Kohn-Sham (KS) ansatz,
which states that the
density can be expressed as a combination of single-particle contributions, one per electron
(or electron pair, depending on whether or not spin is explicitly modeled):
\begin{align}
n \ofr = \sum_{i=1}^{N_\text{e}} | \psi_i \ofr |^2,
\end{align}
where $\psi_i \ofr$ is the KS orbital of the $i$th electron in the system and
$N_\text{e}$ is the number of electrons.
This approximation
allows us to avoid explicitly working with the many-body wave function of the system.
In the KS formulation of DFT, the many-body effects are collected into the exchange-correlation
(XC) density functional $E_\text{xc} [n\ofr]$. The quality of the used approximation for
$E_\text{xc} [n\ofr]$ will determine how close to the actual GS density and energy we can get.

The KS single-particle ansatz, coupled
with the variational principle $\delta E [n] / \delta \psi_i^* = 0$ leads to the eigenvalue-like
KS equation:
\begin{align}
\epsilon_i \psi_i \ofr = H_\text{KS} \psi_i \ofr,
\end{align}
where the KS Hamiltonian $H_\text{KS}$ contains the single-particle kinetic energy operator,
the electrostatic potential and the XC potential. A deeper account of DFT is beyond the scope of
this review, and the reader is referred to the excellent (nowadays almost standard) book
by Martin~\cite{martin_2004} for more detailed information.

The emergence of KS DFT, together with many different approximations to the XC
functional~\cite{perdew_1981,perdew_1996} and efficient
iterative algorithms for solving the KS equation implemented in parallel
computer codes~\cite{kresse_1996} have enabled quantum mechanical calculations of the
properties of matter
at affordable computational cost. In addition to this, the community has been very active at
tackling the different shortcomings of KS DFT, such as the self-interaction error or the lack
of dispersion interactions, e.g., by developing ``hybrid'' XC 
functionals~\cite{,seidl_1996,heyd_2003} and van
der Waals ``corrections''~\cite{grimme_2006,tkatchenko_2009,tkatchenko_2012},
respectively. While DFT is still too expensive to
perform long- and large-scale simulations of atomic systems, the tradeoff between accuracy
and CPU cost afforded by DFT makes it the most popular electronic structure method and,
indeed, the most popular method to generate training data for MLPs.

The purpose of an interatomic potential, also referred to as a force field, is to provide
a computationally affordable approximation of the BO-PES. When training MLPs we often assume
that DFT provides a ``good enough'' version of the BO-PES. While we have just discussed that
DFT has its own shortcomings, it is also important to recognize the limitations of the BO
approximation itself.  Notable breakdowns of the BO approximation occur
whenever protium atoms are present (i.e., the common hydrogen isotope with one proton in
the nucleus)~\cite{habershon_2013}
or when high-energy collisions take place (e.g., during radiation-damage
events in materials)~\cite{darkins_2018}.
Extended MD formalisms are required when simulating these kinds of
systems, for instance time-dependent DFT (TD-DFT) or Ehrenfest dynamics, where electronic
and nuclear degrees of freedom are propagated simultaneously~\cite{casida_2004,li_2005b}. In addition, electronic
excitations and charge-transfer processes~\cite{vlcek_2007} cannot be captured out of the box by
MLPs trained from DFT data. Despite these limitations, which are at the hot spot of
current work by the community, MLPs have enabled incredible successes in computational
materials modeling in recent years, in particular for a-Si and a-C. We review the basics
of MLPs for materials and molecular modeling in the next section.

\subsection{MLP architecture}\label{13}

The rationale for replacing a DFT calculation (or, more generally, an expensive \textit{ab initio}
calculation) by an ML prediction is that, in atomistic systems, the local atomic motifs
are often repetitious. Therefore, put in simple terms, if we compute energies and forces for
a series of reference structures with DFT and store those values in a database, we should in
principle be able to infer a DFT-quality prediction for a new atomic structure as long as
said structure is similar enough to the database entries. The interpolation should be
computationally inexpensive, compared to a DFT calculation, for the procedure to be useful.
The simplest example is a diatomic molecule, where a series of DFT calculations are carried out
for different interatomic separations and a force field is trained from that data to predict
the energy vs distance curve at arbitrary separations. An ``old-fashioned'' empirical
force field would often tackle this problem by fitting the data to a fixed functional
form, e.g., a least-squares fit to a second-order
polynomial. In that sense, MLPs can be regarded as a glorified version of an empirical force
field, where the main difference is the fact that the fit is now carried out in arbitrarily
many dimensions and without the user providing an explicit mathematical function. This is
referred to in ML jargon as a non-parametric fit. Although the distinction between MLPs and
empirical force fields may seem small in this context, in practice the flexibility of ML algorithms
to fit high-dimensional data means that much more complex PESs can be learned, and to much higher
accuracy.

\begin{figure}
    \centering
    \includegraphics[width=\columnwidth]{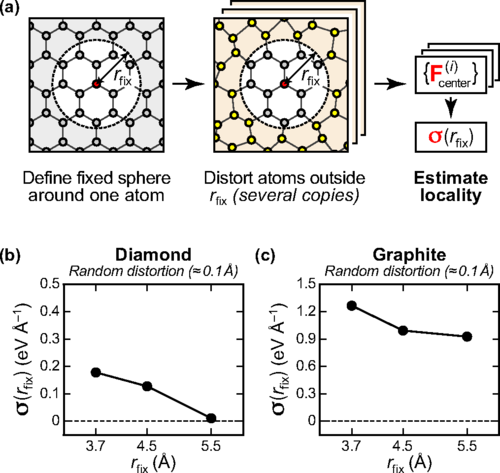}
    \caption{Locality tests in carbon-based systems. (a) Conceptualization of the locality test.
    (b) Convergence of the residual force acting on the central atom for diamond and graphite as a
    function of the cutoff radius. Reprinted from Ref.~\cite{deringer_2017} with permission. Copyright
    (c) 2017 American Physical Society.}
    \label{05}
\end{figure}

Above, two fundamental assumptions are made whose goodness will to a large extent determine
the success of MLPs. One is the assumption of \textbf{locality} of the PES. That is, we can
construct the entire system as a collection of local fragments, each of which has an associated
local energy. Physically, the local energy $\bar{\epsilon}_i$ (where the bar indicates prediction)
is not a well-defined property of the system; instead,
a DFT calculation will return a total energy $E$ for a given ensemble of $N_\text{at}$
interacting atoms. An MLP
will build this total energy from the sum of all the individual contributions which, in simplified
terms, can be considered a sum over atom-wise contributions:
\begin{align}
\bar{E} = \sum\limits_{i=1}^{N_\text{at}} \bar{\epsilon}_i.
\end{align}
An intuitive way to test the locality of the PES for a given material is to monitor the evolution of
the force acting on an atom as other atoms beyond a certain cutoff distance are disturbed, as a function of
said cutoff. This was done in Ref.~\cite{deringer_2017} for crystalline and amorphous C. The procedure
is illustrated in \fig{05}a and the results for diamond and graphite in \fig{05}b are reprinted from
that reference. For diamond (as well as high-density a-C, not shown here but reported in
Ref.~\cite{deringer_2017}) the approximation of locality is extremely
good and the errors are negligible for cutoffs around 5~\AA{} and beyond. For graphite
(and low-density
a-C) the approximation is less good and convergence with the cutoff is very slow. Mathematically,
this approximation implies that we can express a local (atomic) energy prediction as a function of
a finite environment of the atom:
\begin{align}
    \bar{\epsilon}_i = f\left( S_i(r_\text{cut}) \right),
\end{align}
where $S_i(r_\text{cut})$ represents all the relative atomic positions within a sphere of radius
$r_\text{cut}$ centered on atom $i$. In technical terms this means that $\bar{\epsilon}_i$ has
compact support.

The other main assumption is about the \textbf{smoothness} of the PES. That is, a small change in
the positions of the nuclei should lead to a small change in the total energy of the system. In
mathematical terms, the PES should be continuous and continuously differentiable. In a data
science context, smoothness is referred to as regularity.

Besides the training (DFT) data and the two central approximations for the PES, locality and smoothness,
which we have already discussed, an MLP requires also two basic ingredients. The first one is
the atomic structure representation, which is carried out using \textbf{atomic descriptors}. While in
principle the Cartesian coordinates of the nuclei contain all the necessary information, in practice
they are not useful because they do not fulfill the correct symmetries. Specifically, valid atomic
descriptors must fulfill translational, rotational and permutational invariance. The simplest
descriptor is an interatomic distance. More sophisticated descriptors, which contain increasingly
more information about the environment of an atom, can be constructed with a body-order
expansion~\cite{allen_2021}. An interatomic distance is a two-body (2b) descriptor, with a single
degree of freedom.
A 3b descriptor has three degrees of freedom and perfectly characterizes a system made of three
atoms, having subtracted the translation and rotation of the center of mass, which do not
affect energy and forces. Any further body-order increase adds three more degrees of freedom, and
the complexity of the model (and the cost of computing descriptors) explodes with relatively low
body orders. For many practical purposes in materials modeling there is no need to go beyond 3b
terms~\cite{christensen_2020}. However, there is another type of atomic descriptors that allow to
encode the \textit{entire} atomic environment, called many-body (mb) descriptors [cf. \fig{04}c].
Arguably, the most important examples are the smooth overlap of atomic positions
(SOAP)~\cite{bartok_2013} and atom-centered symmetry functions (ACSFs)~\cite{behler_2011}.
It can be shown that these mb descriptors are formally equivalent to one another and, as
constructed from 2b sums within a finite cutoff sphere, are also equivalent to an ensemble
of 3b terms~\cite{willatt_2019}. Two advantages over 3b descriptors are that one mb descriptor
can be used instead of very many 3b ones (since the number of 3b descriptors within a cutoff sphere
explodes as a function of its radius), and that mb descriptors with different numbers of atoms
can be compared to one another (directly relevant in kernel regression methods, cf. \fig{04}e).
The topic of atomic representations is very rich and has been recently summarized in a
comprehensive review paper~\cite{musil_2021}.

The second basic ingredient is the \textbf{machine learning algorithm}. The first method
to interpolate high-dimensional PES with close to DFT accuracy was introduced in 2007 by
Behler and Parrinello~\cite{behler_2007} based on ANNs and applied precisely to model Si. The
second method, based on kernel regression, was introduced by Bart\'ok \textit{et al}. in
2010~\cite{bartok_2010} and used to model C, Si and Ge. Clearly, group-IV semiconductors
have been strongly linked to the use of MLPs since their very inception, and as such
it is unsurprising that the first applications of MLPs to solving outstanding problems
in materials modeling have also focused on C and Si. Naturally, the methodology has advanced
significantly since those two seminal papers and more recent reviews by the authors do
a better job at introducing the concepts and practicalities to the
beginner~\cite{bartok_2015,behler_2017,deringer_2021}. Many other methods and implementations
have appeared since then. A comprehensive account of those is beyond the scope of this work
and so we mention again the recent book summarizing the state of the field~\cite{schutt_2020}.
Below we give a brief overview of these methods, and refer the reader to the cited literature
for further detail.

\textbf{Artificial neural network potential (NNP)}. NNPs~\cite{behler_2007} use artificial
neural networks (ANNs) to interpolate the PES. An ANN consists of a series of ``layers'':
input, hidden and output layers. There is one input and one output layer, and one or more
hidden layers. The input layer contains a vector of features (an ACSF in the case of NNPs)
and the output layer returns an observable, which can be a scalar or a vector (e.g., the
total energy in NNPs). Each hidden layer consists of a number of nodes, and the input data
is propagated forward through the different layers by performing a series of linear and
non-linear operations which depend on the connection and the node in question, respectively.
This propagation procedure is illustrated in \fig{04}d, where the arrows represent the
connections and the circles represent the nodes. We start out with a vector of real-valued
symmetry functions \textbf{G} of a certain dimension, which depends on the number of species
and the quality of the representation~\cite{behler_2011,hellstrom_2020}. Each of these
functions $G_i$ is propagated to each of the nodes in the first hidden layer multiplied
by a series of weights $a^{0,1}_{ij}$ (where {0,1} indicates we are connecting layers 0 and 1):
\begin{align}
\beta^1_j = \sum\limits_{i=1}^{N_0} G_i a^{0,1}_{ij} + b^1_j, \label{27} \\
\gamma^1_j = f (\beta^1_j),
\label{07}
\end{align}
where $N_0$ is the number of nodes in layer 0, i.e., the number of ACSFs (or, equivalently,
the dimension of \textbf{G}) in this case. $b^1_j$ is the bias of node $j$ in layer 1 which,
together with the sum in \eq{27}, define the function $\beta^1_j$, which is linear in the
input variable $G_i$. This quantity, $\beta^1_j$, is used as argument to evaluate a non-linear
\textit{activation function} $f$. The result of this evaluation, $\gamma^1_j$, is then
passed on to the next layer $n = 2$ in the same way as above:
\begin{align}
\beta^n_j = \sum\limits_{i=1}^{N_{n-1}} \gamma_i^{n-1} a^{n-1,n}_{ij} + b^n_j, \\
\gamma^n_j = f (\beta^n_j),
\end{align}
where we note that $N$ can in general vary from layer to layer. We have substituted $G_i$
by $\gamma_i^{n-1}$ for generality, because $G_i$ is the notation used for the input layer
specifically in the case of ACSF for NNPs. This procedure is repeated until
we reach the output layer, which in our case returns a local atomic energy. The forces can be
evaluated analytically from the dependence of the symmetry functions on the atomic positions.

Training an NNP consists in the optimization of the weights $\{ a^{n-1,n}_{ij} \}$ and biases
$\{ b^n_j \}$, and is done using \textit{backpropagation}, for a given training set of atomic
structures, to minimize the error in the corresponding observables (total energies, forces,
stresses, etc.). We will not go into the details of ANN algorithms which, for most practical
purposes in atomistic materials modeling, can be considered a black box.

\textbf{Gaussian approximation potential (GAP)}. GAPs are based on kernel
regression~\cite{bartok_2010} and are arguably more interpretable than ANNs. In GAPs, the local
atomic energy $\bar{\epsilon}_i$ for atom $i$ is expressed as a linear combination of kernel
functions $k$:
\begin{align}
\bar{\epsilon_i} = \delta^2 \sum\limits_{t=1}^{N_\text{train}} \alpha_t k(\textbf{q}_i,
\textbf{q}_t) + e_0,
\label{08}
\end{align}
where $\delta$ is an energy scale, $t$ runs over training configurations, $\alpha_t$ are the
fitting coefficients, $\textbf{q}_i$ and
$\textbf{q}_t$ are the atomic descriptors (often a SOAP mb descriptor)
of a test and train configuration, respectively,
and $e_0$ is a per-atom energy offset, usually taken as the reference energy of an isolated atom
of a given species. The kernel can be understood as a measure of similarity between two atomic
environments, as illustrated in \fig{04}e, and is bounded between 0 (nothing alike) and 1 (identical
up to symmetry operations). Thus, intuitively, the more a training configuration resembles the
test configuration for which we want to make a prediction, the more the fitting coefficient associated
with that training configuration contributes to the prediction. This is why we stated earlier that
GAPs are arguably more interpretable than NNPs.

Having cast the interpolation problem as a linear problem, training a GAP simply consists in a
least-squares-based inversion of \eq{08}:
\begin{align}
\bm{\alpha} = \frac{1}{\delta^2} \textbf{K}^{-1} (\bm{\epsilon} - \textbf{e}_0),
\end{align}
where now the test index in \eq{08} also runs through training configurations, and we do not
use the predicted atomic energy $\bar{\epsilon}$ but the observed one $\epsilon$. We note that in
practice one cannot train a GAP model (or an NNP, for that matter) using local atomic energies,
which are not generally available before training the GAP. Instead,
the local energy in \eq{08} is replaced by the \textit{sum} over local energies leading to a total
energy observable. For instance, when using training data from a DFT calculation for a supercell,
we use $E_I = \sum_i \bar{\epsilon}_i$. In addition to this total energy consideration, one usually
needs to use regularization and sparsification to improve the stability, transferability and
efficiency of a GAP, and may combine several GAPs in the same fit. These details fall outside
the scope of this paper and the reader is referred
to the literature for further insight~\cite{bartok_2015,deringer_2021}. Likewise, the explicit
definition and discussion of atomic descriptors and kernel functions is an active research
topic and better covered
elsewhere~\cite{bartok_2013,caro_2019,willatt_2019,himanen_2020,musil_2021}.
As for NNPs, the forces can be computed analytically through the dependence of $\textbf{q}_i$ on the
atomic positions. Forces and stresses can also be incorporated into the inversion equation, together
with total energies.

\textbf{Other MLP approaches}. The field of ML-based atomistic simulation
of materials is advancing fast. Since NNPs and GAPs appeared, several other MLP flavors have
been developed and we expect applications in amorphous materials modeling to follow soon.
MLP methods besides NNP and GAP include ``linear'' models such as the moment-tensor potential
(MTP)~\cite{shapeev_2016} and the spectral neighbor analysis potential
(SNAP)~\cite{thompson_2015}, or MLPs based on asymptotically complete atomic descriptors
like the atomic cluster expansion (ACE)~\cite{drautz_2019}. The first ACE-based MLP able to
simulate a-C appeared very recently~\cite{qamar_2022}, and it is expected that these
new models and improvements
thereof~\cite{batatia_2022,darby_2022} will overtake NNPs and GAPs as the state-of-the-art
tools for simulating disordered materials in the near future.

In the brief discussion of NNPs and GAPs above we implicitly include short-range interactions only,
since ACSFs and SOAP use radial cutoffs that exclude all interactions beyond a certain radius.
We have therefore left out long-range interactions which are important beyond the
typical cutoffs used to fit ``regular'' MLPs. These long-range interactions include van der Waals
and electrostatics and must be treated on a different footing to bonding and repulsion interactions
both i) out of necessity, to avoid the explosion of computational time with the cutoff distance, and
ii) out of opportunism, since these interactions can often be cast in the form of simple analytical
functions whose \textit{parameters} can be machine learned but are in effect short
ranged~\cite{bereau_2018,veit_2020,xie_2020,muhli_2021b,staacke_2022}.

%\subsection{Comparisons with empirical force fields}

\section{Amorphous and disordered carbon}

\begin{figure}
    \centering
    \includegraphics[width=\columnwidth]{{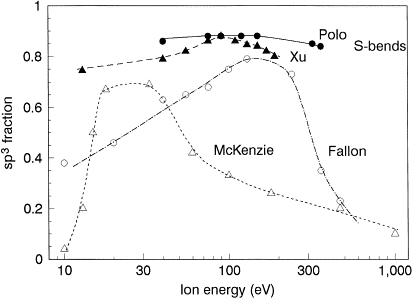}}
    \\[1em]
    \includegraphics[width=0.8\columnwidth]{{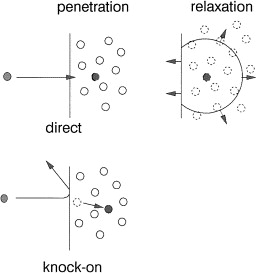}}
    \caption{(Top) $sp^3$ fraction vs deposition energy from a series of literature
    works; data from Polo \etal~\cite{polo_2000}, Xu \etal~\cite{xu_1997},
    Fallon \etal~\cite{fallon_1993} and McKenzie \etal~\cite{mckenzie_1991}.
    The deposition energy is the estimated kinetic energy of individual
    carbon atoms as they hit the substrate. (Bottom)
    The subplantation mechanism postulated as growth mechanism responsible for high
    $sp^3$ fractions in ta-C. Both figures
    reprinted from Robertson~\cite{robertson_2002} with permission. Copyright (c)
    2002 Elsevier.}
    \label{09}
\end{figure}

\begin{figure*}
    \centering
    \includegraphics[width=0.95\textwidth]{{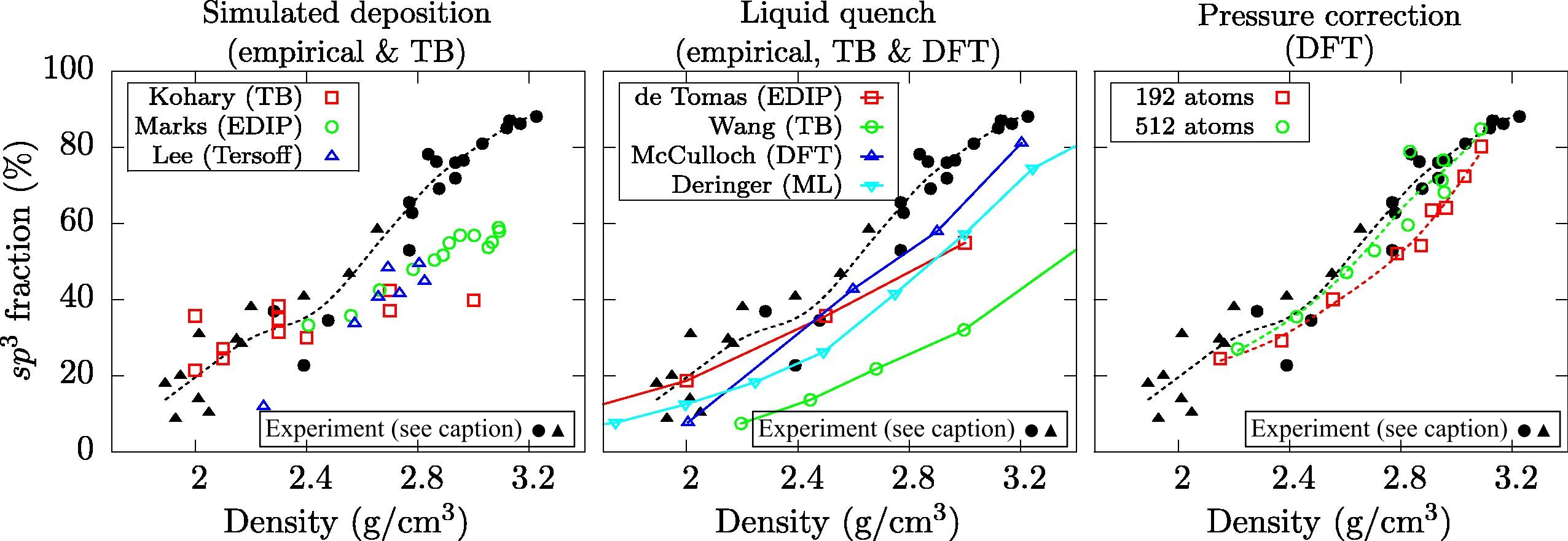}}
    \caption{$sp^3$ fractions vs mass density for different simulation protocols, compared to
    experimental data (black solid dots) from Fallon \etal~\cite{fallon_1993} and (black solid
    triangles) Schwan \etal~\cite{schwan_1996}. Simulation data from Kohary and Kugler~\cite{kohary_2001},
    Marks~\cite{marks_2005}, Lee \etal~\cite{lee_2004}, de Tomas \etal~\cite{detomas_2016},
    Wang and Komvopoulos~\cite{wang_2014}, McCulloch \etal~\cite{mcculloch_2000} and
    Deringer and Cs\'anyi~\cite{deringer_2017}. Pressure-corrected DFT data is taken from
    Caro \etal~\cite{caro_2014} for 192-atom supercells and Laurila \etal~\cite{laurila_2017} for
    512-atom supercells. Reprinted from Ref.~\cite{laurila_2017}.
    }
    \label{10}
\end{figure*}

\begin{figure}
    \centering
    \includegraphics{{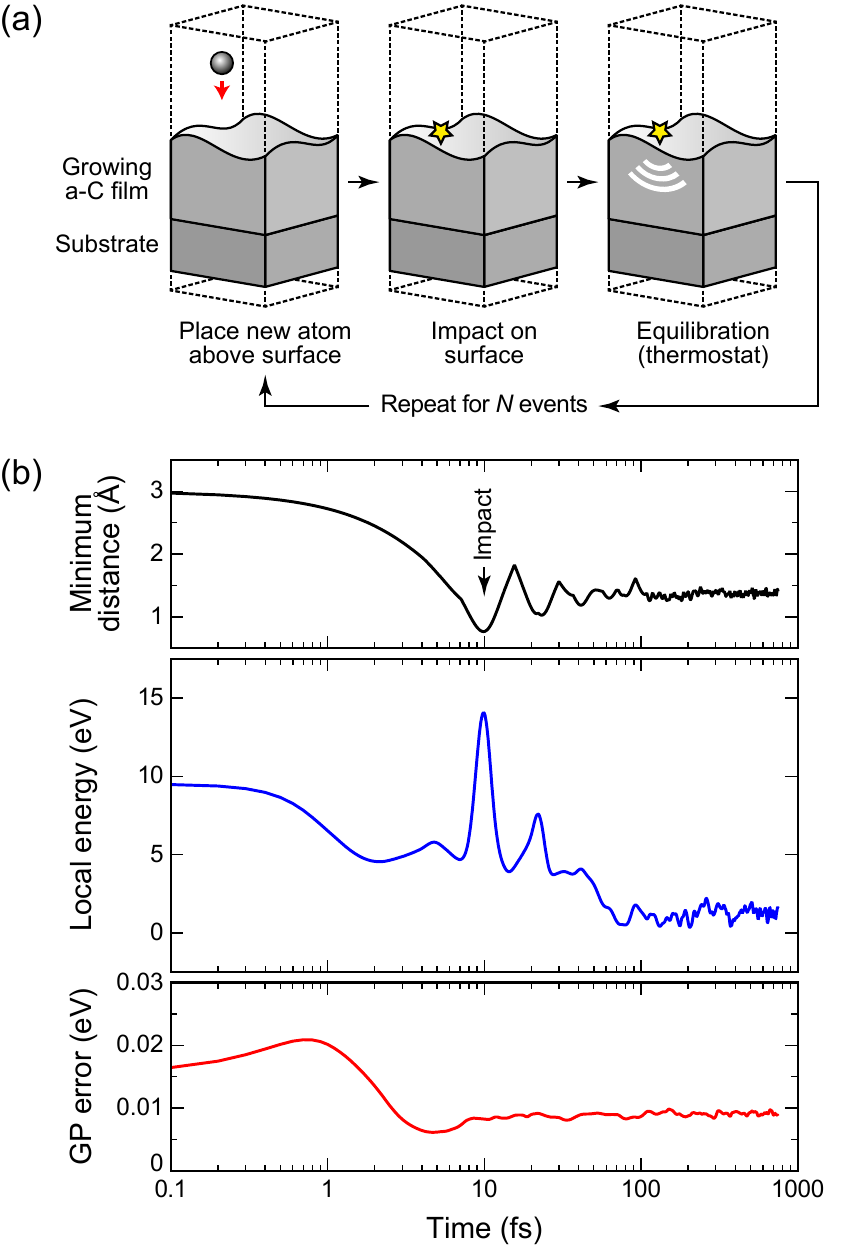}}
    \caption{(a) Schematics of a deposition simulation. (b) Evolution of different
    observables during the course of a single impact event.
    Reprinted from Ref.~\cite{caro_2020c} with permission. Copyright
    (c) 2020 American Physical Society.}
    \label{11}
\end{figure}

\begin{figure*}
    \centering
    \begin{minipage}{0.45\textwidth}
    \begin{flushleft}
    \large{(a)} \vspace{-1em}
    \end{flushleft}
    \includegraphics{{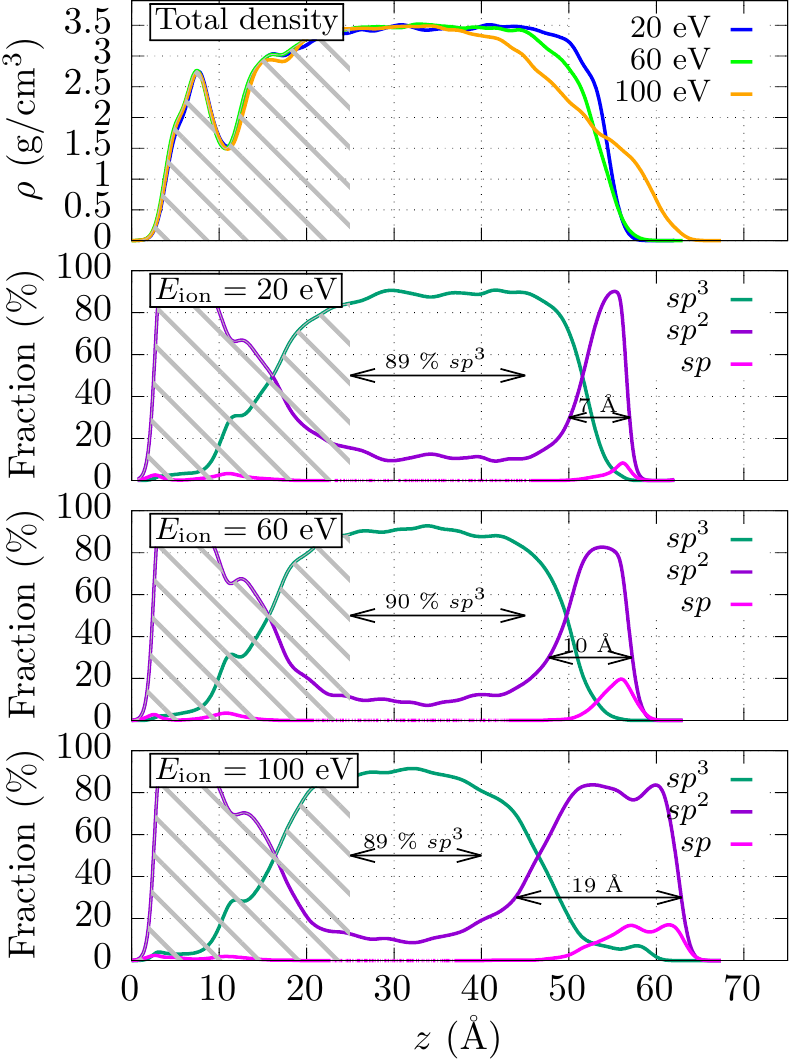}}
    \end{minipage}
    \hspace{0.02\textwidth}
    \begin{minipage}{0.45\textwidth}
    \begin{flushleft}
    \large{(b)} \vspace{-1em}
    \end{flushleft}
    \includegraphics{{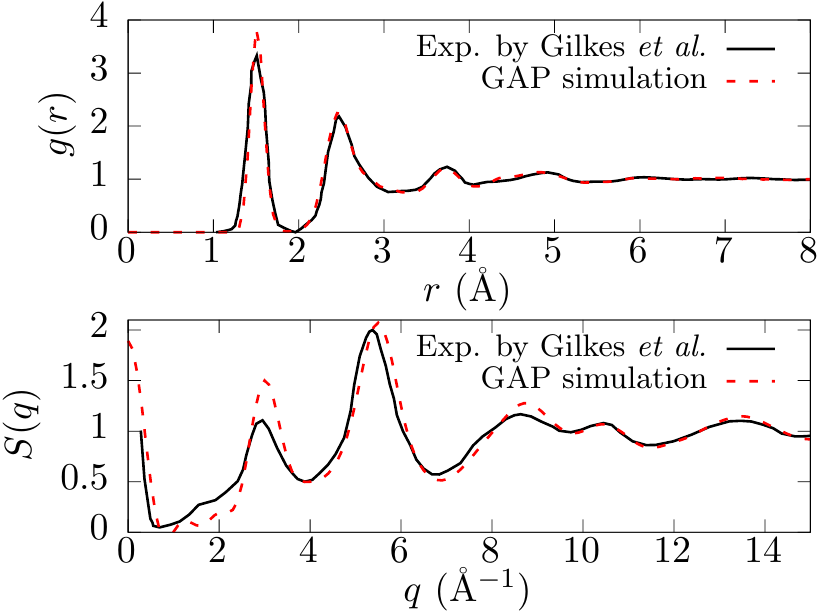}} \\[0.5em]
    \includegraphics{{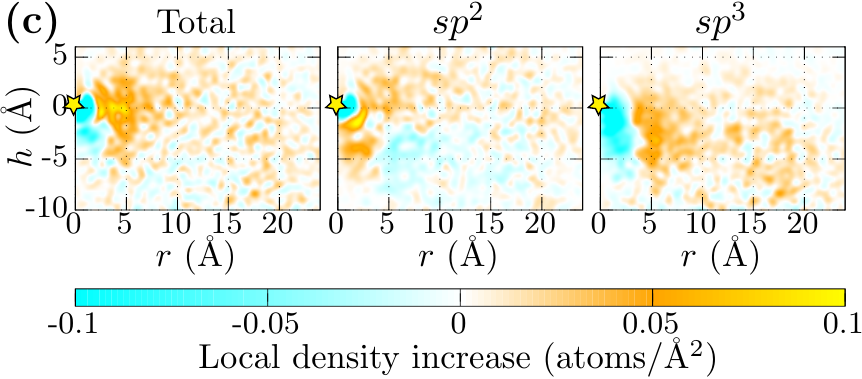}}
    \end{minipage}
    \caption{Results of ta-C deposition simulations:
    (a) Evolution of mass density and coordination fractions along the film's growth
    direction in ta-C over three different deposition energies. The extent of the $sp^2$-rich
    surface, different for each energy, is indicated with an arrow; (b) Radial distribution function
    and structure factor, compared to experimental results from Gilkes \etal~\cite{gilkes_1995};
    (c) Two-dimensional pair-correlation functions indicating the regions of depletion/formation
    of $sp^2$ and $sp^3$ motifs as a function of distance from impact location.
    Reprinted from Ref.~\cite{caro_2018} with permission. Copyright
    (c) 2018 American Physical Society.}
    \label{12}
\end{figure*}

\begin{figure}
    \centering
    \includegraphics[width=\columnwidth]{{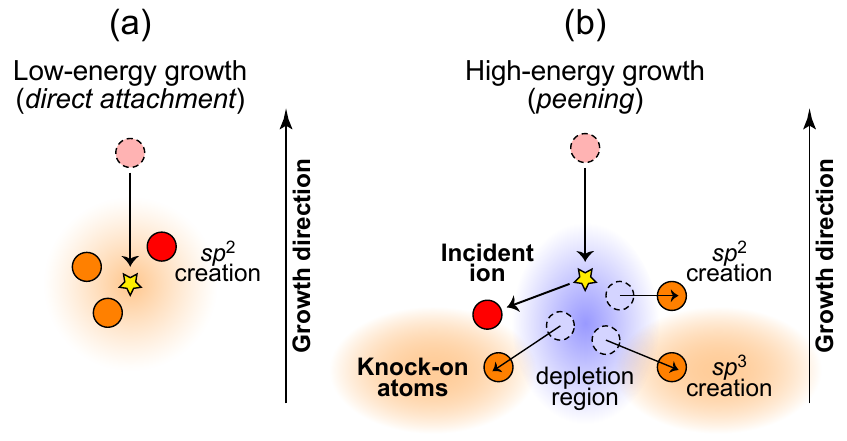}}
    \caption{Observed growth mechanisms in (a) low-density (low deposition energy) and (b)
    high-density (high deposition energy) a-C films.
    Reprinted from Ref.~\cite{caro_2020c} with permission. Copyright
    (c) 2020 American Physical Society.}
    \label{14}
\end{figure}

\begin{figure*}
    \centering
    \includegraphics[width=\textwidth]{{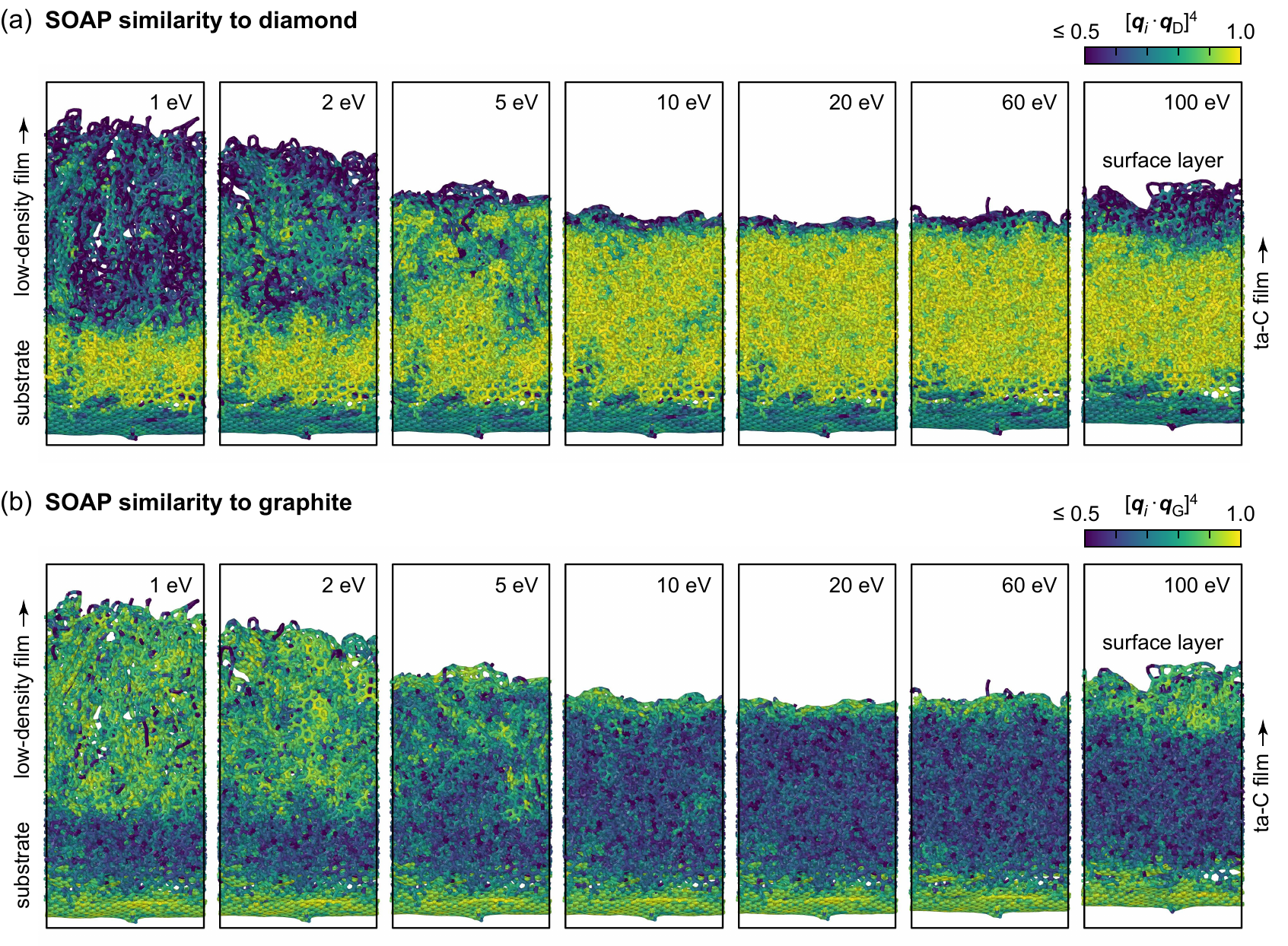}}
    \caption{Evolution of a-C film nanostructure as a function of deposition energy. The top (a)
    and bottom (b) panels show the degree of similarity between atomic environments in the films and
    reference diamond and graphite, respectively. Brighter color indicates more resemblance and
    darker color indicates less resemblance. Reprinted from Ref.~\cite{caro_2020c}
    with permission. Copyright
    (c) 2020 American Physical Society.}
    \label{15}
\end{figure*}

\begin{figure}
    \centering
    \includegraphics[width=\columnwidth]{{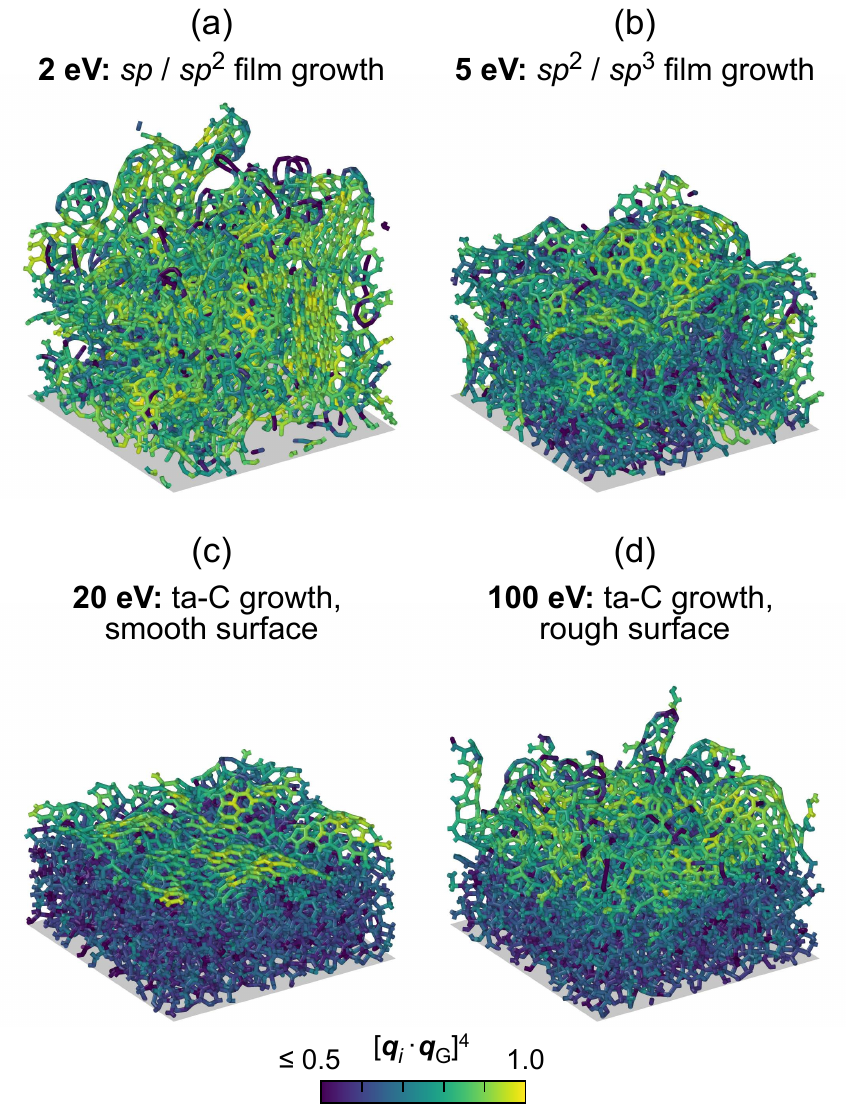}}
    \caption{Detail of the surface nanostructure in a-C films grown at different deposition
    energies, indicating the atomic motifs similarity to graphite.
    Reprinted from Ref.~\cite{caro_2020c} with permission. Copyright
    (c) 2020 American Physical Society.}
    \label{16}
\end{figure}

\begin{figure}
    \centering
    \includegraphics[width=\columnwidth]{{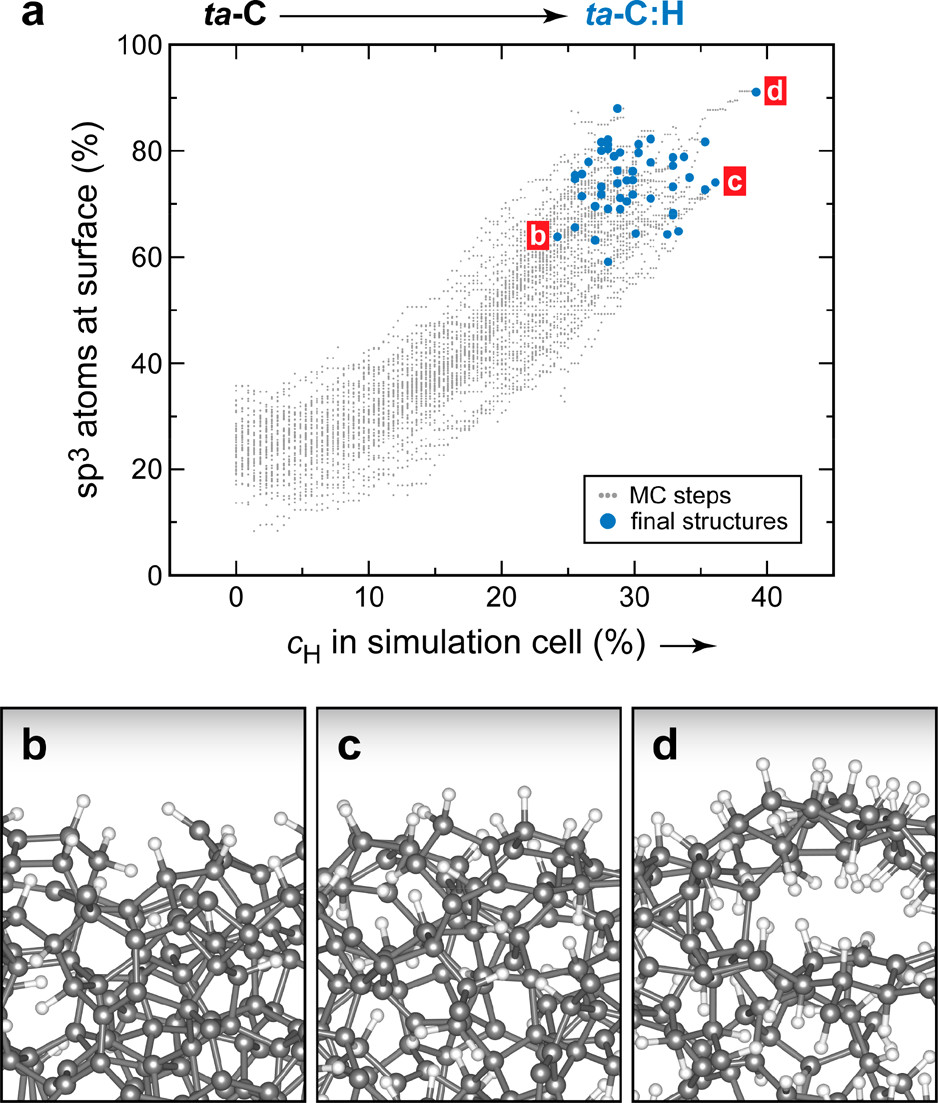}}
    \caption{a) $sp^3$ fraction of hydrogenated a-C as a GCMC simulation progresses where H
    is incorporated. The blue dots mark the end points of the simulation. b-d) Ball-and-stick
    representation of the final a-C:H structures. Reprinted from Ref.~\cite{deringer_2018}
    with permission. Copyright (c) 2018 American Chemical Society.}
    \label{17}
\end{figure}

\begin{figure}
    \centering
    \includegraphics[width=\columnwidth]{{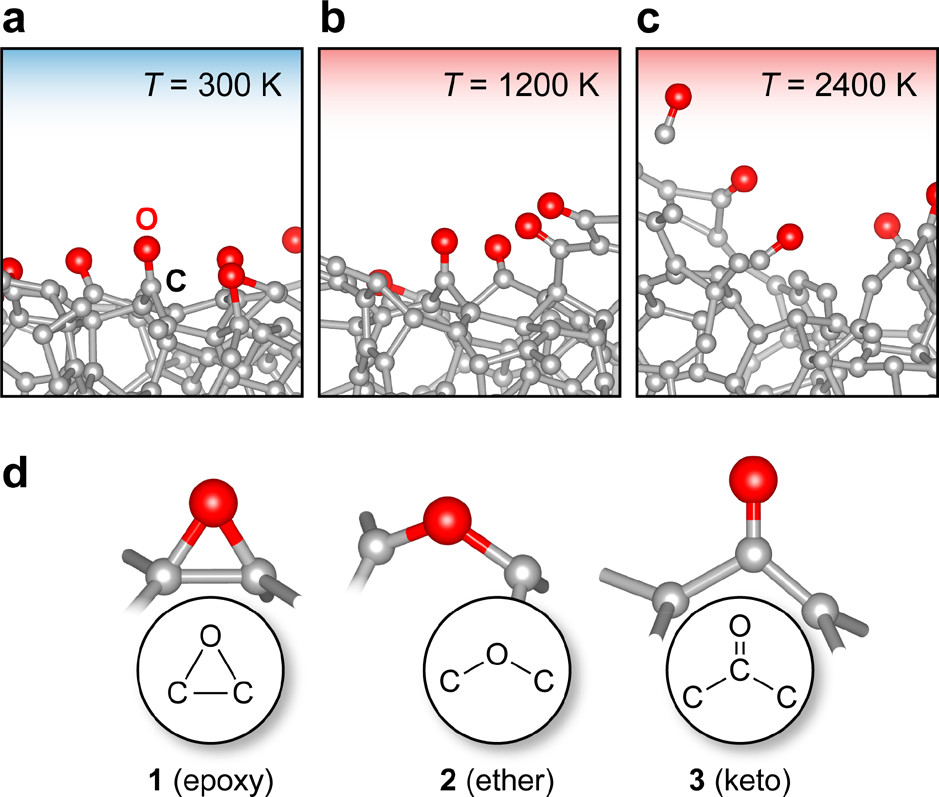}}
    \caption{a-c) Ball-and-stick representation of a-C:O DFT-MD simulations carried out at
    different temperatures. d) Most representative O-containing motifs present in the
    resulting a-C:O structures. Reprinted from Ref.~\cite{deringer_2018} with permission.
    Copyright (c) 2018 American Chemical Society.}
    \label{18}
\end{figure}

\begin{figure*}
    \centering
    \includegraphics[width=\textwidth]{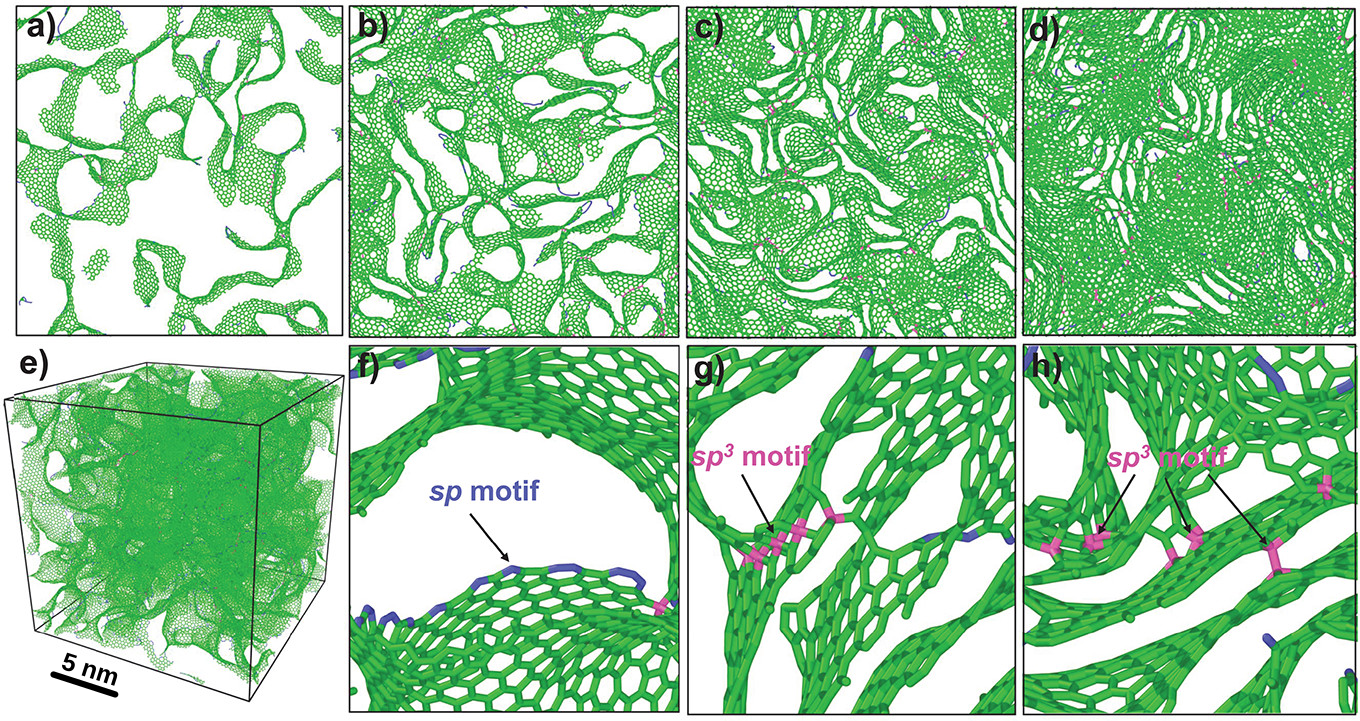}
    \caption{a-d) Nanoporous carbon structures of different densities, where the density
    increases to the right and the average pore size decreases correspondingly.
    e) 3D model of the low-density nanoporous carbon structure from a), where the pore
    morphology can be more easily appreciated. f-h) $sp$ motifs are found in graphitic
    sheet termination (edges) and $sp^3$ motifs are found interlinking stacked graphitic
    layers, conferring three-dimensional rigidity to these nanoporous carbon networks.
    Reprinted from Ref.~\cite{wang_2022}.}
    \label{19}
\end{figure*}

\begin{figure*}
    \centering
    \includegraphics[width=\textwidth]{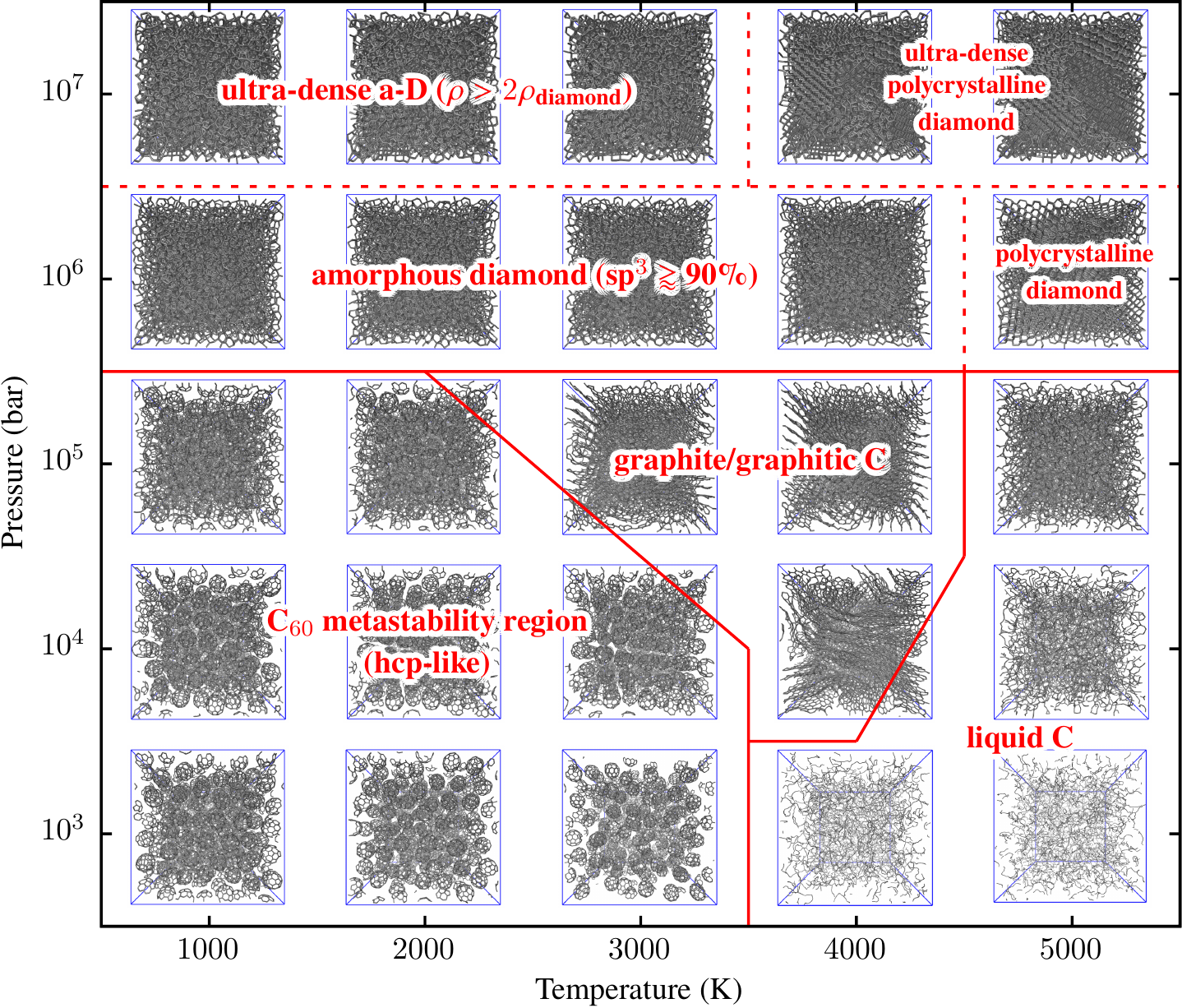}
    \caption{High-pressure/high-temperature phase diagram of C$_{60}$, where
    the molecular precursor leads to nucleation of high-density ta-C (``amorphous
    diamond'') as the molecular C$_{60}$ collapse at high pressure.
    Reprinted from Ref.~\cite{muhli_2021b} with permission. Copyright
    (c) 2021 American Physical Society.}
    \label{20}
\end{figure*}

\begin{figure*}
    \centering
    \includegraphics[width=\textwidth]{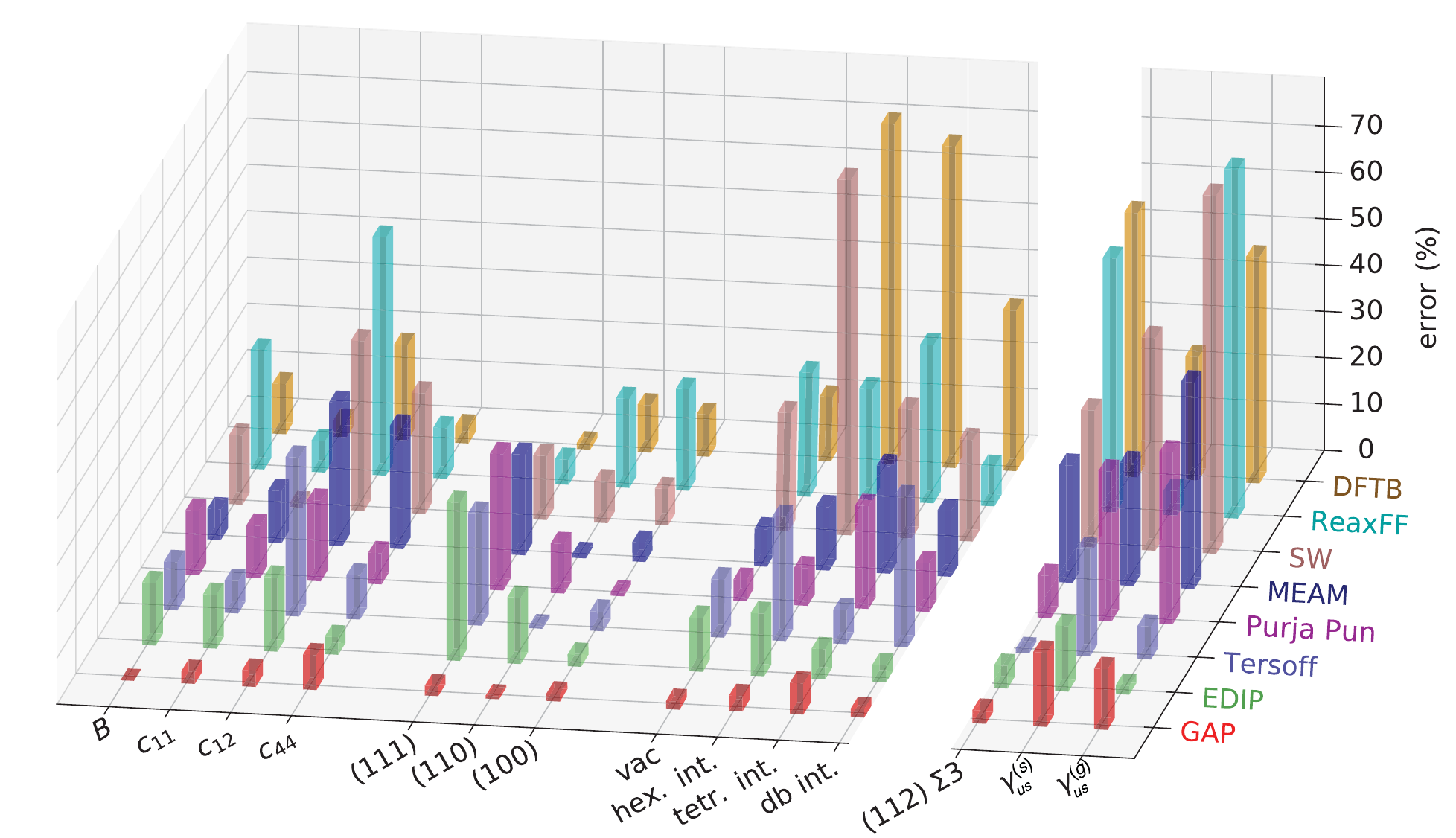}
    \caption{Comparison between the silicon GAP18 and other force fields for predicting
    a number of properties; from left to right: elastic moduli, surface energies,
    point defect formation energies, and planar defect formation energies.
    Reprinted from Ref.~\cite{bartok_2018}.}
    \label{21}
\end{figure*}

\begin{figure*}
    \centering
    \begin{minipage}{0.45\textwidth}
    \begin{flushleft}
    \large{Deringer \textit{et al}.} \vspace{-0.5em}
    \end{flushleft}
    \includegraphics[width=\textwidth]{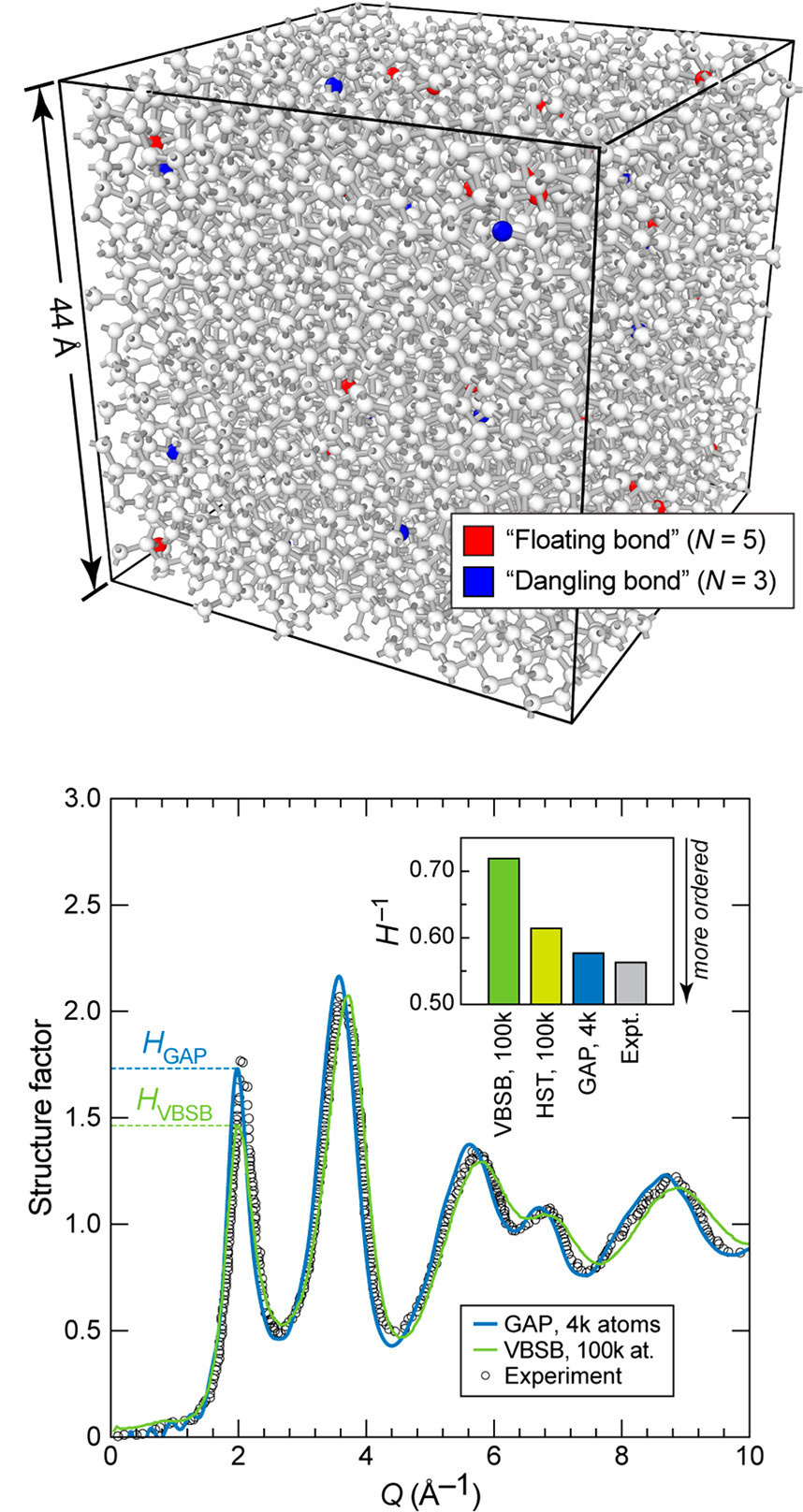}
    \end{minipage}
    \hspace{0.02\textwidth}
    \begin{minipage}{0.45\textwidth}
    \begin{flushleft}
    \large{Wang \textit{et al}.} \vspace{-0.5em}
    \end{flushleft}
    \includegraphics[width=\textwidth]{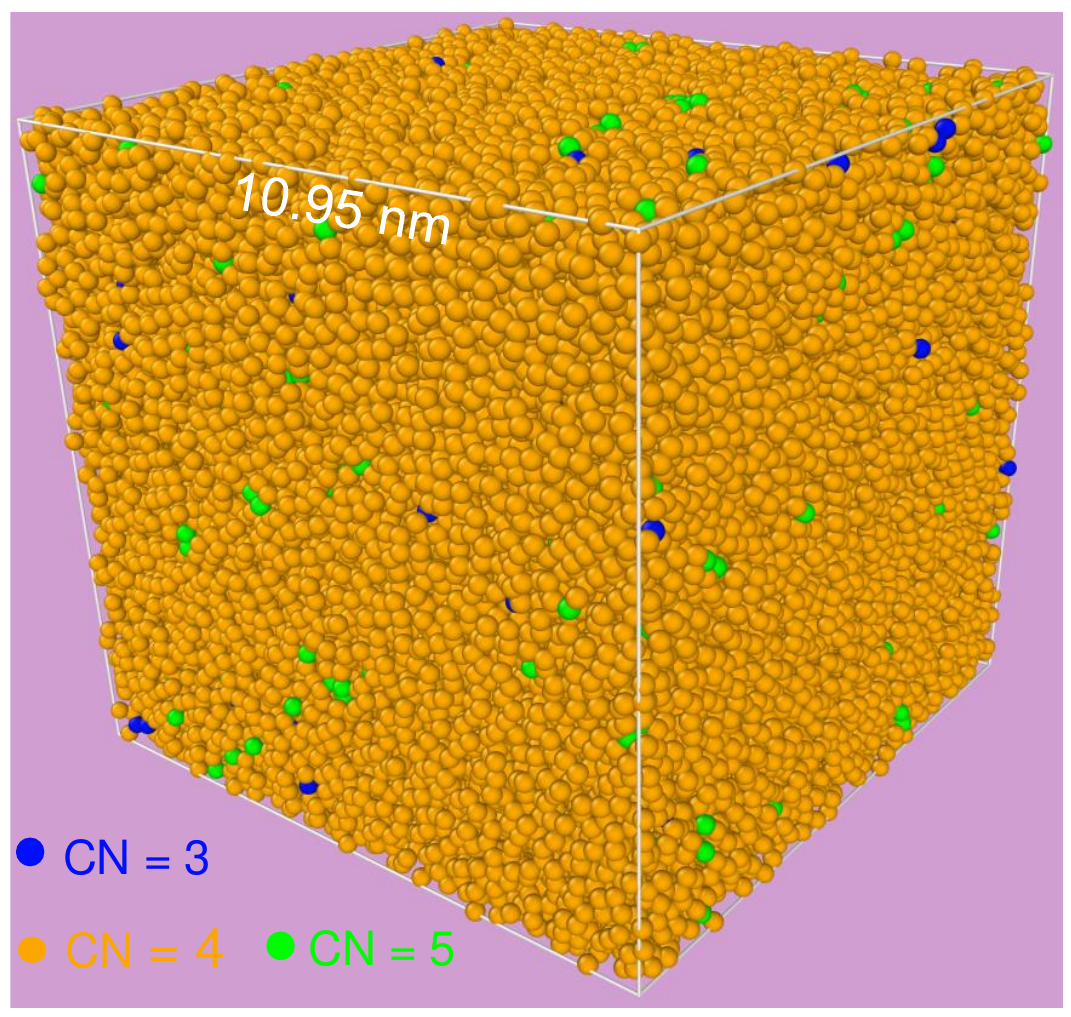} \\[6em]
    \includegraphics[width=\textwidth]{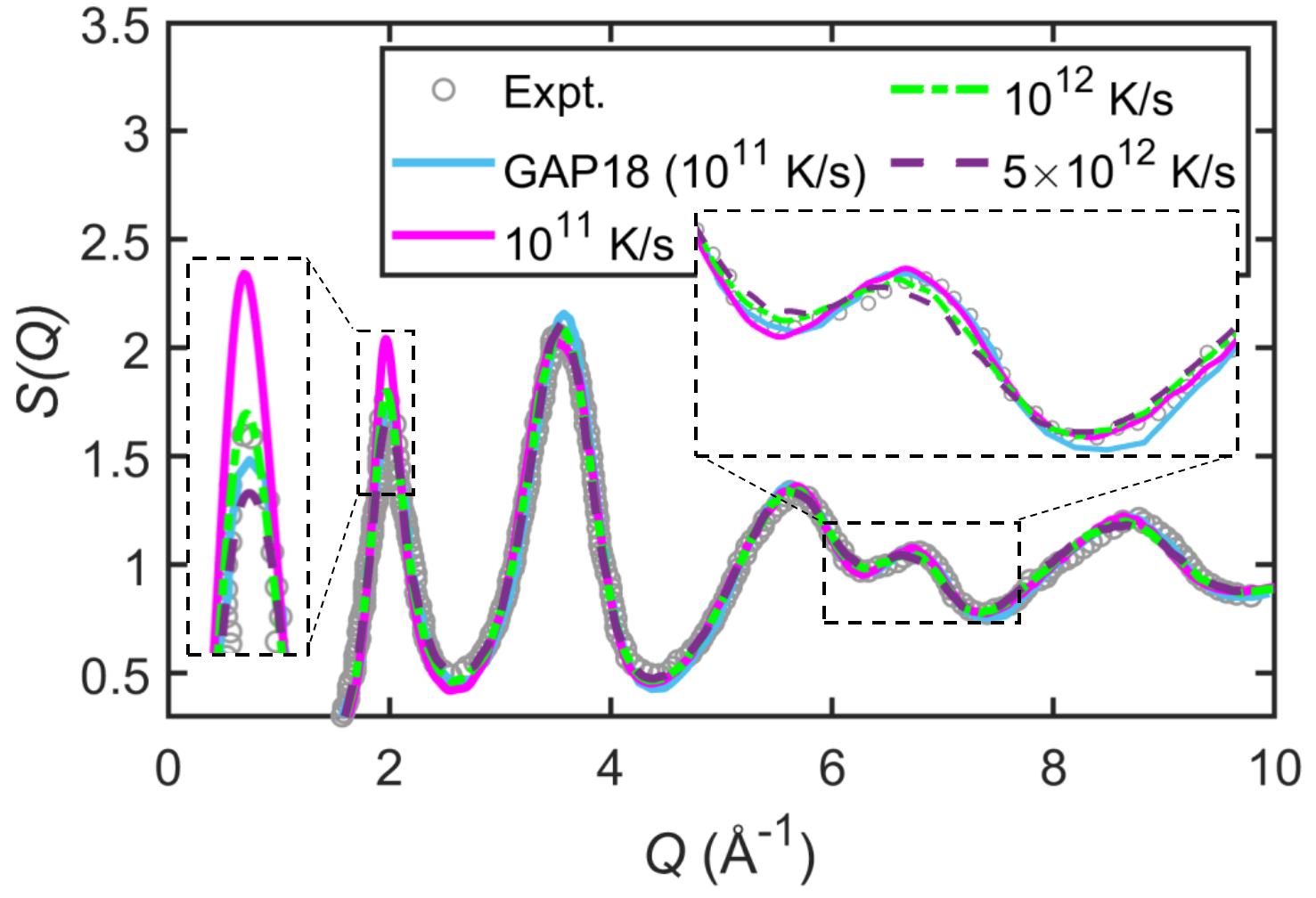}
    \end{minipage}
    \caption{Computational a-Si structural models and their corresponding structure factor
    from Deringer \etal~\cite{deringer_2018b} (left panels) and Wang \etal~\cite{wang_2023}
    (right panels). Deringer's structure factor is compared to that derived from preexisting
    structural models from the literature, VBSB~\cite{vink_2001b} and HTS~\cite{hejna_2013},
    and to experiment. Wang shows the effect of the annealing rate on the structure factor and
    a direct comparison to Deringer's data (labeled GAP18).
    In both cases the experimental data is from Laaziri \etal~\cite{laaziri_1999,laaziri_1999b}.
    Deringer \etal{} figures reprinted from Ref.~\cite{deringer_2018b}. Wang \etal{} figures
    reprinted from Ref.~\cite{wang_2023} with permission. Copyright (c) 2023 American Physical Society.}
    \label{22}
\end{figure*}

\begin{figure}[t]
    \centering
    \includegraphics[width=\columnwidth]{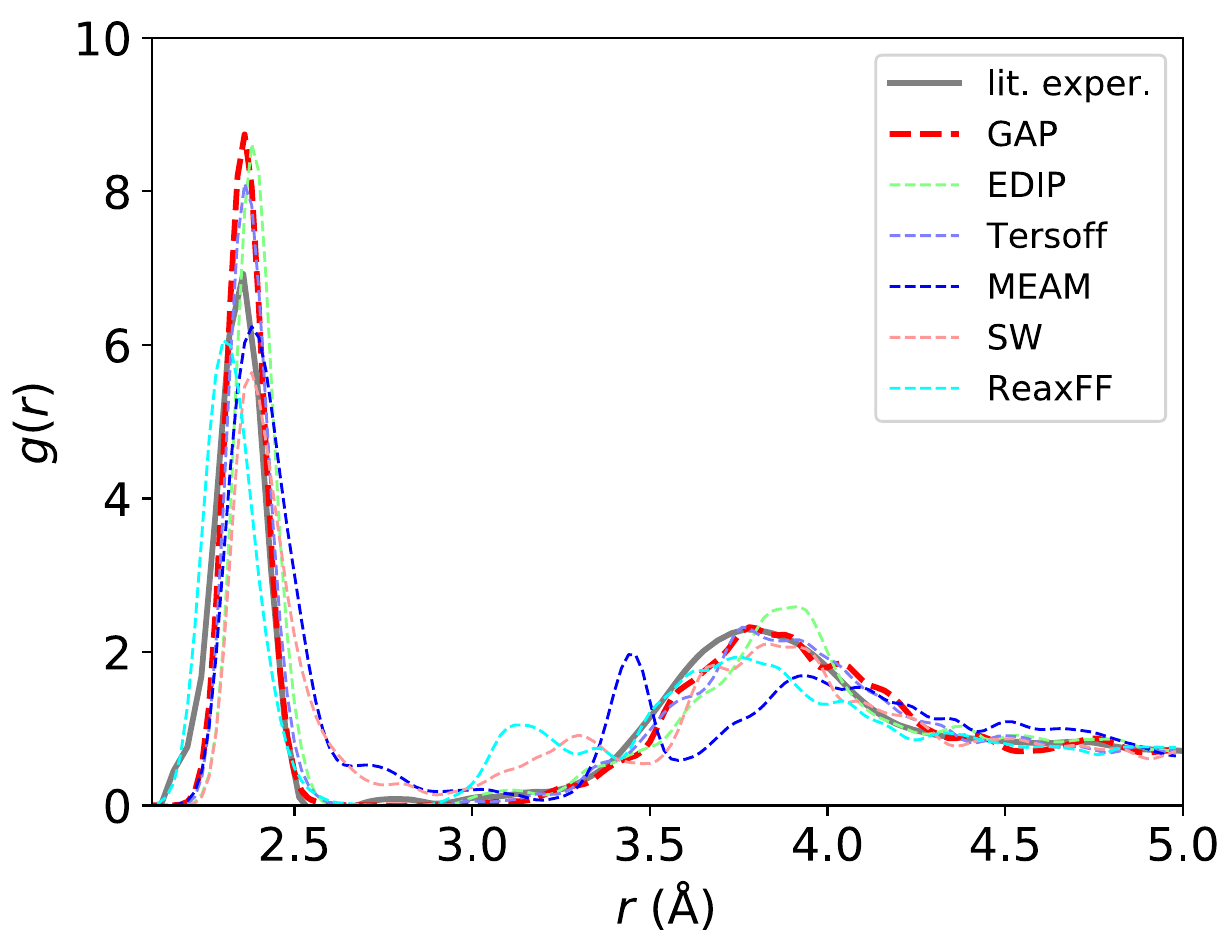}
    \caption{GAP results for the RDF of a-Si from Bart\'ok \etal~\cite{bartok_2018},
    and comparison to the RDF obtained from experiment~\cite{laaziri_1999,laaziri_1999b}
    and using different popular interatomic force fields for silicon.
    Reprinted from Ref.~\cite{bartok_2018}.}
    \label{23}
\end{figure}

\begin{figure}[t]
    \centering
    \includegraphics[width=\columnwidth]{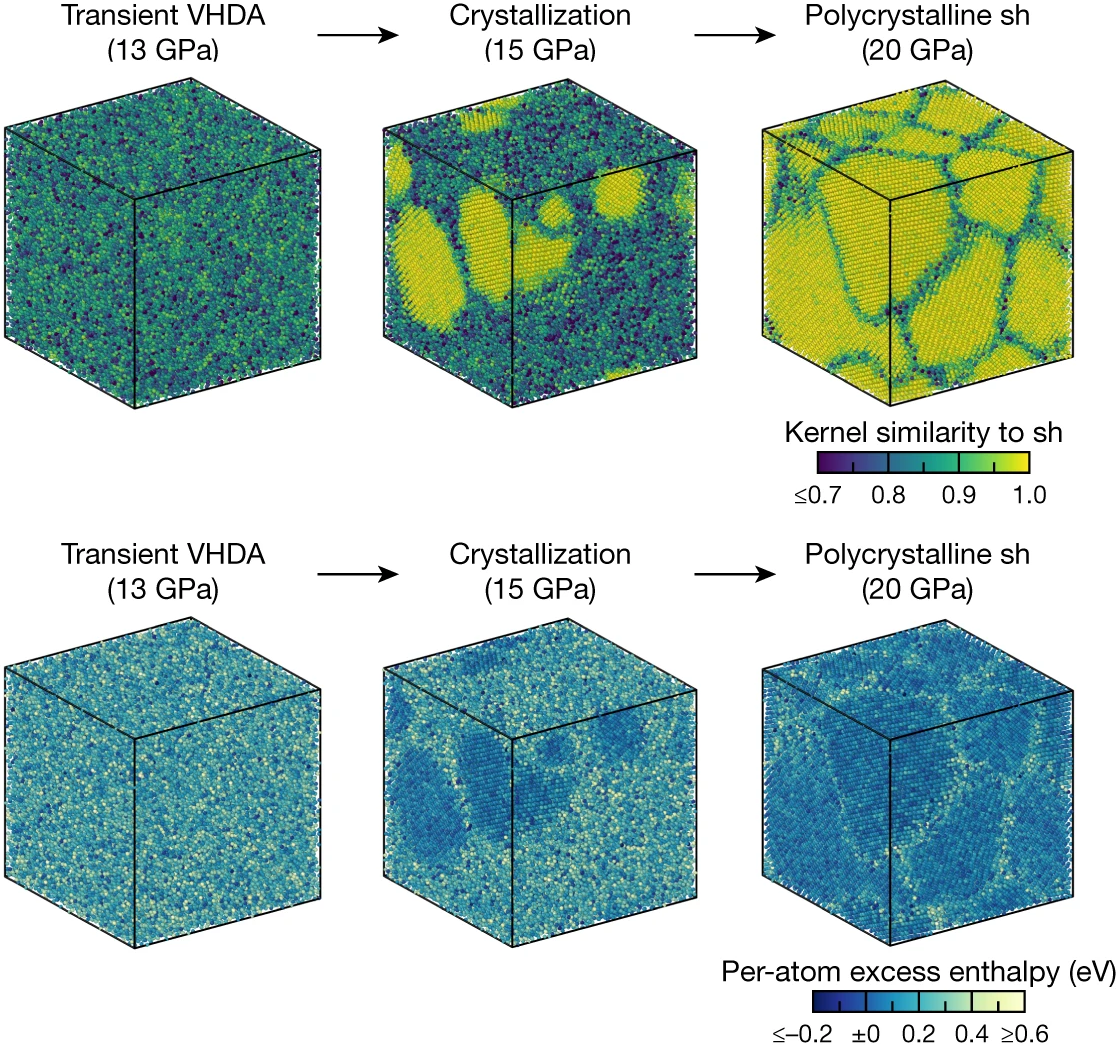}
    \caption{High-pressure phase transition in silicon from a very-high-density amorphous (VHDA)
    phase to a (metallic) polycrystalline phase. Reprinted from Ref.~\cite{deringer_2021b}
    with permission. Copyright (c) 2021 Springer Nature.}
    \label{24}
\end{figure}

The precise structure of a-C and how growth conditions can be tuned to modify
it have been the topic of intense debate for several decades. The reason is that, unlike in a-Si,
in a-C coordination environments with different number of atomic neighbors, ranging from two
to four neighbors ($sp$ and $sp^3$ orbital hybridizations, respectively) are all possible
(meta)stable motifs. Especially three- ($sp^2$) and four-coordinated environments can coexist in
a-C thin films. Varying the relative concentration of $sp^2$ and $sp^3$ bonding in a-C allows us to
tune its material properties (mechanical, electrical, optical, etc.) from graphite-like to
diamond-like. Thus, much of the basic characterization work on a-C has dealt with the dependence
of the $sp^2$/$sp^3$ ratio on such parameters as the deposition energy during physical vapor
deposition (PVD) growth and, in turn, the dependence of the material properties on the
$sp^2$/$sp^3$ ratio. The reference entry point into the properties of a-C, albeit a bit
outdated in terms of missing atomistic simulation insights that were developed in the last
few years, is Robertson's monumental review paper from 2002~\cite{robertson_2002}. Two of the most
important figures in that paper are reprinted here in \fig{09}. The top panel shows the dependence
of the $sp^2$/$sp^3$ ratio on the deposition energy (the estimated kinetic energy of incident
C atoms) for different experimental techniques used to grow a-C. A common trend is the increase
in $sp^3$ content for increasing deposition energy, up to around 100~eV, after which there is
a further decline. We also note that some of these a-C samples can be grown with extremely
high $sp^3$ contents, approaching that of diamond (100~\% $sp^3$). Therefore, diamondlike or
``tetrahedral'' a-C (DLC and ta-C, respectively) can be made experimentally, offering a route
for cheap coatings with diamondlike hardness for tribological applications. The bottom panel
in \fig{09} shows the proposed film growth mechanism in ta-C, subplantation, which was widely regarded
as the correct mechanism until recently, when MLP simulations~\cite{caro_2018} provided
quantitative confirmation for earlier qualitative evidence~\cite{marks_2005} of an alternative
mechanism, peening, which we will discuss later in more detail.

Thus, our journey into simulation of a-C starts with the extensive atomistic modeling efforts
carried out in the pursuit of explaining how these high $sp^3$ contents can be achieved and,
in turn, explaining the growth mechanism in a-C. The state of the art in atomistic modeling
of a-C, as of 2017 (just 6 years before this review was completed), using a variety of
interatomic potentials (including DFT) and modeling
techniques, is summarized in \fig{10}. This figure was compiled right after the first MLP able
to model a-C was published, also in 2017~\cite{deringer_2017}. We can see that the direct simulation
route, deposition, fell short of achieving the very large $sp^3$ contents observed experimentally
at high deposition energies. This technique had, back in the day, been restricted to fast empirical
methods, such as tight binding (TB), C-EDIP and Tersoff potentials, and was (and still is) computationally
unfeasible at the DFT level. Liquid quench simulations, on the other hand, were accessible to more
expensive methods, like DFT, but could only predict very large $sp^3$ contents at unphysically high
pressures, which shows as a shift towards higher mass densities on the graph. On that middle
panel of \fig{10} we note the first ML-based simulations of a-C generation with the MLP trained
by Deringer and Cs\'anyi~\cite{deringer_2017}, a stepping stone in a-C modeling and key development
leading to the advances that we will discuss later. Finally, pressure-corrected DFT simulations
from Refs.~\cite{caro_2014,laurila_2017}, based on a two-step relaxation procedure,
managed to get extremely good agreement with experiment
for the $sp^3$ content as a function of mass density but, based on an indirect simulation protocol,
offered no insight whatsoever into the growth mechanism.

\textbf{Explaining the growth mechanism}. With the introduction in 2017 of the first MLP
able to handle the structural complexity of a-C with
close to DFT accuracy, but at a fraction of the computational cost, the first MLP-based simulation
depositions followed soon. In 2018, Caro \etal~\cite{caro_2018} presented MD deposition
simulations of ta-C growth. In these simulations, incident atoms with varying kinetic energy
(20, 60 and 100~eV) impinge on a carbon substrate, initially diamond. After each impact, the
system is equilibrated back to the nominal deposition temperature (300~K) and the next deposition
event takes place. After several thousands of atoms have been deposited, the size of the film is
enough to collect statistics for material properties and growth mechanism. The workflow of these
deposition simulations is shown in \fig{11}(a). Panel (b) of the figure shows a more detailed view
of the impact process for a single event with a logarithmic time axis. Initially, the incident atom
approaches the substrate very fast. The MLP algorithm allows us to monitor the local atomic energy,
as we have discussed in Sec.~\ref{13}, shown in the figure offset by the average local energy in the
growing film. The highly energetic initial impact is followed by equilibration of the atomic environment,
where atoms settle in their new positions. GAP MLPs also allow us to monitor the predicted interpolation
error, shown in the figure too. This deposition process is rather complex, with the order of 50
bond breaking/formation events taking place for each impact at around 100~eV~\cite{caro_2020c}.
The process is better visualized as a video animation, with several Open Access resources
available from the
literature, including a single impact~\cite{caro_2020d}, the atom-by-atom growth of a-C thin films
from low to high density~\cite{caro_2017d}, and the resulting atomic structures in XYZ
format~\cite{caro_2020} (which enable subsequent studies).

Figure~\ref{12} shows the key results from these first deposition simulations~\cite{caro_2018}.
Panel (a) shows the mass density
and $sp$/$sp^2$/$sp^3$ fraction profiles along the growth direction for the deposition energy ranges
where ta-C growth takes place. The three simulations at 20, 60 and 100~eV result in similar mass
densities and coordination fractions in the bulk of the film, for the first time close to those reported
experimentally for the densest ta-C films. At the same time, the data shows rather different surface
morphologies, with increasing surface roughness for higher deposition energies, a result that
closely follows experiment~\cite{davis_1998}. This, together with the also excellent agreement with
experiment for the radial distribution function (RDF) shown in \fig{12}(b) gave confidence in the
quality of the simulations as representative of the microscopic growth mechanism taking place
experimentally. The collection of deposition statistics (up to 8000 individual events per energy)
then enabled drawing a precise picture of what that growth mechanism actually looks like. In
\fig{12}(c) we can see the mass density and coordination fraction increase/decrease before and
after an impact event, averaged over all impacts, as a function of depth and lateral separation
from the impact site. The color maps clearly indicate a local decrease of $sp^2$ and, especially,
$sp^3$ carbons around the site of impact. In fact, these maps show that locally (around the impact
site) there is an increase in the amount of $sp^2$ carbon, whereas the $sp^3$ fraction increases
laterally and away from the impact site, due to pressure waves originating from the impact region.
This mechanism is known as peening, and had already
been proposed by Marks in 2005 on the basis of deposition simulations with the C-EDIP empirical
potential~\cite{marks_2005}. Because the C-EDIP simulations lacked quantitative agreement with
experiment [cf. \fig{10}], the peening mechanism did not gather generalized adoption. However,
the GAP deposition simulations offer strong quantitative support for peening as the growth mechanism
in ta-C, which is schematically illustrated in \fig{14}(b).

\textbf{a-C structure across mass densities}. Following the study of ta-C growth, these
deposition simulations were extended to low-density
a-C and a more comprehensive analysis of material properties and comparison with other interatomic
potentials was carried out~\cite{caro_2020c}. At low deposition energies (below the typical
cohesive energy per atom in carbon materials, $\lessapprox 9$~eV), graphitic a-C grows by direct
attachment [\fig{14}(a)], where higher coordination increases the stability of $sp$ surface
motifs by creating $sp^2$ and $sp^3$ carbon. At very low deposition energies of a couple of eV
the rate of $sp^3$ formation is small and the bulk of the a-C films is highly graphitic, with
$\sim 80$~\% $sp^2$ and similar amounts of $sp$ and $sp^3$ motifs ($\sim 10$~\% each). Figure~\ref{15}
shows the evolution of the structure, as well as the ``graphite likeness'' and ``diamond likeness'',
of the simulated thin films as a function of deposition energy. The similarity to graphite and
diamond is computed by calculating the SOAP kernels between each individual atomic environment
and either pristine graphite or diamond~\cite{deringer_2017,caro_2020c}. Since the kernel can be
understood as a similarity measure, bounded between 0 and 1, as we have previously discussed, it
can also be used for this kind of quantitative comparison in a very straightforward way.
At and above $\sim 10$~eV the
structure is largely homogeneous and dominated by $sp^3$ motifs in the bulk of the film. At 5~eV
$sp^2$ and $sp^3$ motifs are approximately equally frequent, and the material shows a ``patched''
structure. At low deposition energies the material is graphitic in nature, as we have already discussed,
and made of tubular (nanotube-like) structures and highly defective graphitic sheets.

\textbf{a-C surface structure}. The surface structure in a-C is always graphitic like (Figs.~\ref{12}
and \ref{15}), but the extent of this $sp^2$-rich surface layer is strongly dependent on the deposition
energy. The MLP deposition simulations show a smoothest surface at around 20~eV and roughest at 100~eV
(and, presumably, higher energies, which have not been studied)~\cite{caro_2018}. For very low deposition
energies there is no well-defined surface region, except for a higher relative abundance of $sp$ motifs
within the top 10~\AA{} or so, and the film is graphitic and highly disordered throughout. A closeup
on the structure of a-C surfaces for different mass densities, again color coding according to
graphite likeness, is given in \fig{16}.

\textbf{Doped a-C}. As-deposited a-C is rarely completely free of impurities. Residual elements
present in the reactor chamber and sample setup lead to unintentional doping of a-C films with
a wide range of chemical species, the most significant of which are H, O, N and
Si (see, e.g., time-of-flight elastic recoil detection analysis [ToF-ERDA]~\cite{mizohata_thesis}
results of the elemental makeup of a-C:N~\cite{etula_2021}).
Besides unintentional doping, it is possible to incorporate impurities in order to achieve
a desired effect. The mechanical, electronic and (electro)chemical properties of a-C can be
modified via H, O and N incorporation~\cite{robertson_2002,santini_2015,etula_2021}.
Intentionally doped a-C is usually denoted by a-C:X,
where X stands for the dopant. The most common doped form of a-C is hydrogenated a-C,
or a-C:H, where typical H contents are of the order of
30--50~at.-\%~\cite{robertson_2002}.
This material can be made for instance by depositing C in a H/methane plasma, or by depositing
acetylene or methane molecules directly, depending
on the desired C/H ratio~\cite{robertson_2002}.
The properties of a-C:H differ from those of a-C in that
H atoms will saturate many bonds, potentially leading to high $sp^3$ contents but relatively low
mass densities, because a H atom is about 12 times as light as a C atom. The mechanical properties
of a-C:H, e.g., its elastic moduli, will be inferior to those of ta-C with similar $sp^3$
content~\cite{robertson_2002}.

Computational studies on a-C:H are comparatively rare, and direct deposition simulations of a-C:H
with MLPs are not yet available in the literature (although our group is currently exploring this
possibility). Indirect simulations of a-C:H formation and H adsorption
energetics mixing MLPs and DFT or other electronic-structure methods have appeared in recent
years. Ref.~\cite{deringer_2018} did grand-canonical Monte Carlo (GCMC) simulations of H adsorption
using density-functional tight binding from a wide set of preexisting a-C surface models. These
results showed that a-C:H materials with very high $sp^3$ fractions ($\sim$~75\%) can be obtained
over a relatively large range of H concentrations (ranging from 25\% to 40\%). The results of these
GCMC trajectories and snapshots of the resulting films are given in \fig{17}.
Ref.~\cite{caro_2018c} focused on the individual adsorption processes and how those depend on the
geometry of the preexisting pure carbon sites, finding two separate regions of adsorption stability
for $sp$ and $sp^2$ sites each. Almost-linear and almost-planar $sp$ and $sp^2$ motifs,
respectively, are more stable and therefore less reactive towards H adsorption than more bendy
motifs. Ref.~\cite{caro_2018c} also introduced a series of ML models reminiscent of MLPs to accurately
predict adsorption energies. These models are based on SOAP descriptors and were augmented with
electronic structure information, directly incorporating the local density of states
into the structural descriptor, which allowed to significantly increase the model accuracy.

Growth of a-C:O and a-C:N usually takes place by introducing O$_2$ and N$_2$ into the deposition
chamber~\cite{santini_2015,etula_2021}. As with a-C:H, the amounts of dopant
that can be incorporated is significantly higher than in traditional doped semiconductors,
with maximum values of at least 30~at.-\%~\cite{santini_2015,golze_2022} and
10~at.-\%~\cite{etula_2021} for O and N incorporation, respectively,
reported in the literature.
By adjusting the partial gas pressure the concentration of dopants can be adjusted.
Again, Refs.~\cite{deringer_2018} and \cite{caro_2018c} appear to be the only published work where
MLPs were used to study O incorporation into a-C, although this was done less comprehensively
than for H. In particular, Ref.~\cite{deringer_2018} did DFT-based MD simulations of a-C surface
oxidation starting from a MLP-generated surface model. The resulting structures are shown in
\fig{18}. This work established the predominant oxygen-containing motif in a-C:O surfaces
being keto-like groups. ML models have also been recently used to
understand the structure of H- and O-containing disordered carbon materials through atomistic
simulation of X-ray photoelectron spectroscopy (XPS)~\cite{golze_2022}. XPS and other spectroscopies
are expected to provide increasingly stronger links between experiment and simulation
as ML techniques for
atomistic structural characterization continue to evolve.

We are not aware of atomistic studies of a-C:N based on MLPs, although our group is currently
developing an MLP able to handle the CN system over a wide range of structures, including a-C:N.
We are also developing MLPs for the CH and CO systems, which will hopefully shed light onto the
structure and properties of a-C:H and a-C:O. Our more ambitious objective in the longer term is to
combine these into a CHO(N) MLP able to accurately describe a wide variety of carbon-based materials
and molecules under different thermodynamic conditions. This objective will likely be achieved,
either by us or by others, within the next couple of years.

\textbf{Nanoporous carbon}. Nanoporous (NP) carbon is related to low-density (highly graphitic)
a-C. They share the overall graphitic nature of their chemical bonds and the lack of long-range
order, but NP carbons are organized in less defective graphitic layers with very low $sp$ or
$sp^3$ content. The usefulness of NP carbons resides in their porous structure and how it can
be exploited in particular for ion intercalation in energy-storage solutions~\cite{wang_2021},
such as Li-ion or Na-ion batteries and supercapacitors. GAP-derived structural models
have already been used to understand intercalation mechanisms and diffusion in NP carbon
materials~\cite{lahrar_2019,lahrar_2021}.
Graphitic carbons
of different densities can be generated computationally following
a ``graphitization'' protocol. This is a special
kind of melt-quench simulation where there is a long annealing step at the graphitization
temperature~\cite{detomas_2016} which, for GAP MLPs, is around 3500~K~\cite{detomas_2019}.
MLPs have been used to study the intercalation of Li and other alcali-metal ions in graphitic
carbons using small-scale structural models~\cite{fujikake_2018,huang_2019}.
More recently, GAP simulations by Wang \etal~\cite{wang_2022} have produced high-quality
large-scale (> 130,000 atoms) structural models of NP carbon throughout a wide range of mass
densities, some of which are shown in \fig{19}. In these materials, the relative abundance of
5- and 7-ring defects (the stable ring motif in graphite is a 6-ring) determine the
curvature of the graphitic planes and thus the pore morphology. There are slightly more 5-rings
than 7-rings in these materials, and nanopore sizes and morphologies seem to be
rather homogeneous for a given mass density, according to the results of this study. The
mechanical properties of the materials were found to evolve smoothly with density. We note that
these NP structural models are already one order of magnitude bigger than those also obtained
with MLPs just a couple of years prior, highlighting the rapid pace of development in the field.

\textbf{Disordered carbon under extreme conditions}. Atomistic simulation is a particularly
attractive approach to study matter under conditions that make direct experimentation complicated
or even impossible. This is the case for high-temperature and high-pressure conditions under
which some allotropes are stable (notably, diamond is stable at very high pressures). The landscape
of carbon allotropes is particularly rich given the flexibility of carbon covalent bonding.
Traditional search strategies for new crystals can be accelerated by using MLPs which are able to
navigate the PES with close to \textit{ab initio} accuracy but orders of magnitude faster. New
carbon allotropes have been found following this approach~\cite{deringer_2017b}. MLPs also
enable us to go one step further thanks to the increased computational efficiency and chart phase
transformations for disordered materials explicitly (i.e., beyond the small unit cells used
for crystal structure search). An example of this is the large-scale study of the phase diagram
of C$_{60}$ carried out by Muhli \etal~\cite{muhli_2021b}. In this work, phase transformations
from a C$_{60}$ precursor at high temperatures and pressures were simulated with an MLP, successfully
leading to the prediction of a transformation to amorphous diamond (a-D) from the collapsed
precursor, later observed experimentally at similar thermodynamic conditions~\cite{shang_2021}.
The detailed phase diagram, shown in \fig{20}, required structural models with thousands of atoms
to correctly describe the configurational disorder. Furthermore, this work exemplifies the power
of MLP simulation, where accuracy can be maintained within a unified methodological framework
across a wide range of thermodynamic conditions,
from low pressures and temperatures where weakly bonded (e.g., van der Waals) interactions dominate
to the extreme conditions at which a material phase collapses into another. This highlights
the potential of MLPs to chart unknown phases of materials taking different precursors as starting
point and applying a range of physical transformations on them, which is of particular importance
for the discovery of new carbon materials.

\section{Amorphous silicon}

Just as they have opened up new avenues in carbon simulation, MLPs have also enabled computational
atomistic studies of silicon that were out of reach just a few years ago. Silicon is the
archetypical semiconductor and, for this reason, the two seminal papers on atomistic ML for
materials modeling used Si as a proof-of-concept material~\cite{behler_2007,bartok_2010}.
Indeed, Behler and Parrinello even looked at the MLP-predicted structure of liquid Si and
compared it, favorably, to DFT results~\cite{behler_2007}. Possibly, this was the first-ever
MLP simulation of a ``disordered'' material. Much has happened in MLP modeling of a-Si in the
few years since those seminal papers appeared, which is summarized in this section.

\textbf{General-purpose Si MLPs}. The first prerequisite on the way to accurate atomistic
simulation of an amorphous material is the availability of a general-purpose potential. The
first MLP of this type for silicon was introduced in 2018 (the year following the introduction of the
first general-purpose carbon MLP) by Bart\'ok \etal~\cite{bartok_2018}. The authors lucidly proposed
a materials-property benchmark as a more relevant accuracy test for their potential than the
prevalent train/test splits. These comprehensive tests are summarized in \fig{21}, showing how
this GAP MLP outperforms a wide selection of other potentials available at the time of the
comparison. The tested properties include elastic moduli, surface energies, point defects,
and planar defects. Further tests not shown in the figure included bulk crystal properties,
liquid and a-Si RDFs, phase diagram, phonons, and more, including a simulation of crack
propagation. These tests give confidence in the quality and broad applicability of the MLP
to diverse problems, an important requirement for modeling the complex structure of a-Si with
high accuracy, especially to obtain the correct concentration of coordination defects,
both under- (3-fold) and over-coordinated (5-fold) motifs. This Si GAP
was extensively validated specifically for a-Si simulation in Ref.~\cite{deringer_2018b}.
The database of structures generated
by Bart\'ok \etal{} has been used to train new versions of the MLP~\cite{caro_2021b,wang_2023},
and their MLP has enabled important subsequent work elucidating the atomic structure of a-Si,
summarized below.

\textbf{The atomic structure of a-Si}. Compared to a-C, the structure of a-Si may seem
relatively simple since every motif which is not made up of a 4-fold coordinated atom is a
coordination defect. However, the defect concentration can have a massive impact on the
optoelectronic properties of this material and, therefore, a force field's success at
modeling a-Si resides in being able to capture these subtleties. In particular, the
predicted relative formation energies for over (5-fold) and under (3-fold) coordinated defects
under varying local strain field will determine the quality of the computational structural
models that can be generated. In \fig{21} we see how the GAP MLP from Ref.~\cite{bartok_2018},
let us call it GAP18 for short,
outperforms all other available force fields in simultaneously predicting the correct
elasticity and defect energetics in silicon. This basis provides the confidence required to
trust the a-Si structures derived from melt-quench simulations. This confidence is reinforced by
comparing the structure factor of simulated a-Si with experimental results. Figure~\ref{22}
shows results from Refs.~\cite{deringer_2018b} and \cite{wang_2023} using GAP18~\cite{bartok_2018}
and a NN MLP (``neuroevolution potential'', NEP~\cite{fan_2021}) trained from the GAP18 database,
respectively. Both works derive similar levels of coordination defects, 0.5\% 3-fold and 1\%
5-fold defects from Ref.~\cite{deringer_2018b}, and 
0.45\% 3-fold and 1.45\% 5-fold defects from Ref.~\cite{wang_2023}. These MLP simulations
show very good agreement with experimental structure factors over a wide range of wavelengths,
free of the artifacts encountered by other simulation methods, as shown in \fig{22}(left-bottom)
and \fig{23} (for the RDF, which has information equivalent to that contained in the structure
factor).

The medium-range order, often quantified in terms of ring counts~\cite{franzblau_1991},
shows prevalence of the stable 6-membered ring motif (slightly less than one per atom, on average)
followed by relatively large amounts of defect ring motifs: 7-membered ($\lesssim 0.6$ per atom),
5-membered ($\lesssim 0.4$ per atom) and 8-membered ($\lesssim 0.1$ per atom), followed by almost
negligible amounts of larger and smaller rings~\cite{deringer_2018b,wang_2023}.
Comparing to carbon, a-C has significantly broader ring distributions at low density but
similar ones at high density~\cite{caro_2014},
whereas nanoporous carbon has significantly narrower distributions centered on 6-membered rings
and skewed towards 5-membered rings, instead of towards 7-membered rings~\cite{wang_2022}.

The most sophisticated study, to date, on the structure and structural transitions in disordered Si
have been recently presented in the hallmark study by Deringer \etal{}~\cite{deringer_2021b}. The
authors used large-scale structural models with up to 100k atoms and long-time-scale MD simulations
to reproduce the liquid-to-amorphous phase transition at high temperature (around 1180~K). This
transition is characterized by a rapid decrease in the number of over-coordinated
($N_\text{neighbors} > 4$) structural motifs and rapid increase in the resemblance of \textit{local}
a-Si atomic motifs to those in crystalline Si, even in the absence of mid- or long-range order.
An even more impressive part of this work is the characterization of a highly non-trivial high-pressure
phase transition between the amorphous semiconducting phase and the (poly)crystalline metallic phase
via intermediate nucleation of crystallites embedded within the amorphous matrix (\fig{24}). The metallic
nature of the high-pressure crystalline phase was established with a previously developed ML model for
the electronic density of states (DOS)~\cite{benmahmoud_2020}. Indeed, integration of ML interatomic potentials
with other ML-based approaches that feed on similar or the same descriptors, for example
charge partitioning for model parametrization~\cite{muhli_2021b} or core-level energies for
X-ray spectroscopy~\cite{golze_2022}, is opening the door for ever more sophisticated simulation
of the properties of disordered materials.

\begin{figure}[t]
    \centering
    \includegraphics[width=\columnwidth]{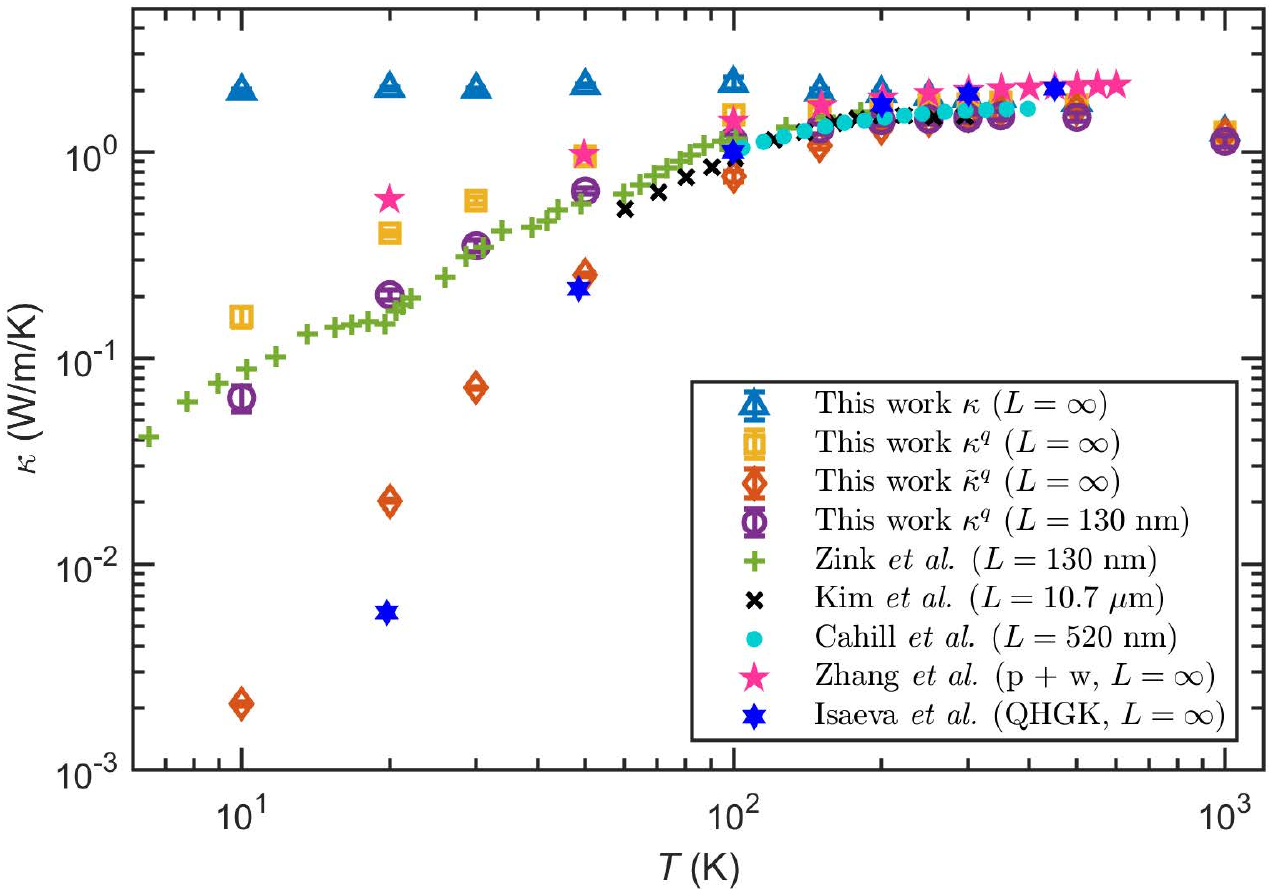}
    \caption{Evolution with temperature of the thermal conductivity of a-Si; comparison between
    simulation and experiment. ``This work'' here refers to Ref.~\cite{wang_2023}. Experimental
    references are to Zink \etal{}~\cite{zink_2006}, Kim \etal{}~\cite{kim_2021},
    Cahill \etal{}~\cite{cahill_1994}, Zhang \etal{}~\cite{zhang_2022} and Isaeva
    \etal{}~\cite{isaeva_2019}. Reprinted from Ref.~\cite{wang_2023} with permission. Copyright (c) 2023 American Physical Society.}
    \label{25}
\end{figure}

\textbf{Properties of a-Si}. While insight into the structure of a-Si is
interesting on its own, for device applications we are also interested in emerging material properties,
such as electronic band gap or thermal transport mechanism. To predict and understand these properties
the key lies in finding the link between them and the underlying atomistic structure of the material.
The number of MLP-driven studies of these properties in a-Si still lags behind the more developed
literature on its atomistic structure, for the simple reason that the existence of reliable structural
models precedes the calculation of properties, which must necessarily rely of the availability of those
models. We expect to see rapid development in characterization and prediction of the properties of
a-Si within large-scale atomistic simulation in the next few years. We mention here a recent example
on thermal conductivity in a-Si. Wang \etal{}~\cite{wang_2023} used an ANN-based NEP 
MLP~\cite{fan_2021}, implemented in the GPUMD code~\cite{fan_2022},
to perform highly efficient homogeneous non-equilibrium MD simulations of thermal transport in a-Si.
This study has been able to closely reproduce the experimental evolution of the thermal conductivity
of a-Si as a function of temperature and finite size (\fig{25}). This opens the door,
in the near future, to simulating
thermal properties of materials yet to be synthesized, with a high level of confidence in the
accuracy of the computational results, in turn providing the basis for accelerated materials discovery
and property-based materials design.

\begin{table}
\caption{List of some of the publicly available MLPs to simulate disordered C and Si, together with
practical information. ``Code'' refers to the computer software which can be used to run a
simulation with the corresponding MLP. GP stands for ``general purpose'', i.e., it refers to
an MLP which is not tailored exclusively to simulate a disordered phase but can be reliably
used for that purpose. Whenever more than one version of the MLP exists, we give the reference
to the latest one (i.e., on the table v2 refers to version 2 of a given MLP).}
\begin{ruledtabular}
\begin{tabular}{l | l | l | l | p{2.5cm}}
Material & Year & MLP flavor & Refs. & Code(s)
\\ \hline
a-C & 2017 & GAP & \cite{deringer_2017} & QUIP, LAMMPS
\\
C (GP) & 2020 & SNAP & \cite{willman_2020} & LAMMPS
\\
a-C & 2021 (v2) & GAP & \cite{caro_2020b,wang_2022} & QUIP, LAMMPS, TurboGAP
\\
C (GP) & 2021 & GAP & \cite{muhli_2021,muhli_2021b} & QUIP,\footnotemark[1] LAMMPS,\footnotemark[1] TurboGAP
\\
C (GP) & 2022 (v2) & GAP & \cite{rowe_2020,csanyi_2022} & QUIP, LAMMPS
\\
a-C & 2022 & NEP & \cite{fan_2022,ref_nep_pots} & GPUMD
\\
C (GP) & 2022 & ACE & \cite{qamar_2022} & LAMMPS
\\ \hline
Si (GP) & 2018 & GAP & \cite{bartok_2018,csanyi_2021} & QUIP, LAMMPS
\\
Si (GP) & 2021 & GAP & \cite{caro_2021b} & QUIP, LAMMPS, TurboGAP
\\
Si (GP) & 2021 & NEP & \cite{fan_2021,ref_nep_pots} & GPUMD
\\
a-Si:H & 2022 & GAP & \cite{unruh_2022} & QUIP, LAMMPS
\end{tabular}
\end{ruledtabular}
\footnotetext[1]{Full van der Waals corrections for this MLP are only available with TurboGAP.}
\label{26}
\end{table}

\textbf{Hydrogenated a-Si}. While a-Si is an interesting material from a fundamental scientific 
perspective, as the prototypical amorphous semiconductor, hydrogenated a-Si (a-Si:H) is arguably
more important from a technological point of view. a-Si:H is commonly used in solar cells, where the
intentional doping with H heals the coordination defects that are present in undoped a-Si, improving
material properties towards photovoltaic applications. Unfortunately, the introduction of additional
chemical species makes the development of accurate MLPs more challenging, because of the larger
configuration spaced spanned. At the same time, the CPU cost of an MLP calculation with multiple
species is more expensive (``curse of dimensionality'') than for single species, with descriptor
construction typically scaling between exponentially (worst-case scenario) and linearly
(best-case scenario) with the number of species. For this reason there are comparatively few studies on
a-Si:H or Si alloys compared to pure Si. On the other hand, thanks to recent developments in MLP
technology, such as descriptor compression~\cite{darby_2022}, and the improved collective
expertise gained by the community on how to generate good databases to train transferable MLPs, these
multispecies force fields are starting to emerge. Here we mention in particular the recent effort by
Unruh \etal{} to develop an a-Si:H GAP~\cite{unruh_2022}. The new Si-H GAP shows quantitative agreement
with DFT and a significant improvement upon previously available classical (non-ML) force fields. This
new MLP enabled the authors to model nanopores in a-Si:H. In the near future, either this force field
or an extension combining its training database with existing and new ones may enable device-size
simulation of c-Si/a-Si:H heterojuctions towards mitigating degradation mechanisms~\cite{jordan_2017},
of major technological importance for high-efficiency solar cells.

\section{Available MLPs}

Table~\ref{26} provides a non-comprehensive list of available MLPs able to simulate a-C and a-Si,
together with their ML ``flavor'' and which code(s) they can be used with. These potentials are
mostly of the GAP flavor, since the GAP community has been the most active at simulating a-C and a-Si
among the different MLP developers and users base.

\section{Summary and outlook}

In this Topical Review we have introduced MLPs as powerful tools for the simulation of
disordered materials at the atomic scale, making it possible to accurately study atomistic
systems within sizes and time scales that were out of reach just a few years ago. We have
discussed how these new tools have been used to shed light on important questions for
understanding the structure of a-C and a-Si. MLPs have been used to elucidate the growth
mechanism in diamondlike carbon and the high-pressure phase transformation from a-Si to
a high-coordination metallic Si phase. MLPs have been used to study phase transitions at
extreme thermodynamic conditions in carbon materials, and to understand the structure of
nanoporous carbon, a material of increasing importance in battery research. These tools
are being extended to more complicated systems, in particular H- and O-doped a-C and
H-doped a-Si, enabling a further degree of realism in simulating these structurally complex
materials. Furthermore, MLPs are being coupled to other ML approaches, including
electronic-structure and spectroscopic signature prediction, improving the prospects for
direct comparison and better integration between experiment and simulation. All in all,
the future of atomistic
modeling of disordered materials, also beyond a-C and a-Si, looks bright in the wake of
MLPs. We should expect important breakthroughs in materials research in the years to come,
brought about by these new powerful computational tools.

\begin{acknowledgments}
The author is grateful to Prof. Volker L. Deringer from the University of Oxford
for useful comments on this manuscript, and to the Academy of Finland for personal financial support,
under Research Fellow grant \#330488.
\end{acknowledgments}

\end{document}